\documentclass[12pt,letterpaper]{article}
\pdfoutput=1 

\usepackage{amsmath,amssymb,calc}
\usepackage{graphicx}
\usepackage[nosort]{cite}
\usepackage{epsfig,psfrag}

\makeatletter
\renewcommand\section{\@startsection {section}{1}{\z@}%
                                   {-3.5ex \@plus -1ex \@minus -.2ex}
                                   {2.3ex \@plus.2ex}%
                                   {\normalfont\large\bfseries}}
\renewcommand\subsection{\@startsection{subsection}{2}{\z@}%
                                     {-3.25ex\@plus -1ex \@minus -.2ex}%
                                     {1.5ex \@plus .2ex}%
                                     {\normalfont\bfseries}}
\makeatother

\let\non\nonumber

\let\a=\alpha
\let\b=\beta

\let\s=\sigma

\let\Th=\Theta

\newcommand{\del}{\partial}
\newcommand{\delbar}{\bar{\partial}}

\def\one{^{(1)}}

\newcommand{\bea}{\begin{eqnarray}}
\newcommand{\eea}{\end{eqnarray}}
\newcommand{\be}{\begin{equation}}
\newcommand{\ee}{\end{equation}}
\newcommand{\bma}{\begin{pmatrix}}
\newcommand{\ema}{\end{pmatrix}}

\newcommand{\mat}[2]{{\left(\begin{array}{#1} #2 \end{array}\right)}}

\newcommand{\hlf}{\frac{1}{2}}

\newcommand{\Z}{{\mathbb Z}}

\newcommand{\F}{{\mathbb F}}
\newcommand{\PP}{{\mathbb P}}
\newcommand{\CC}{{\mathbb C}}

\newcommand{\cC}{{\cal C}}
\newcommand{\cA}{{\cal A}}
\newcommand{\cF}{{\cal F}}
\newcommand{\cJ}{{\cal J}}

\renewcommand{\O}{\operatorname{O}}

\newcommand{\La}{\Lambda}
\newcommand{\la}{\lambda}

\newcommand{\G}{\Gamma}
\newcommand{\e}{\epsilon}

\newcommand{\dd}{\delta}

\newcommand{\m}{\mu}
\newcommand{\n}{\nu}
\newcommand{\p}{\partial}

\newcommand{\f}{\psi}

\newcommand{\tf}{\widetilde{\f}}

\newcommand{\ap}{\alpha'}

\newcommand{\D}[1]{\ensuremath{\mathrm{D}#1}}

\newcommand{\C}[1]{$(\ref{#1})$}

\def\IZ{\relax\ifmmode\mathchoice
{\hbox{\cmss Z\kern-.4em Z}}{\hbox{\cmss Z\kern-.4em Z}}
{\lower.9pt\hbox{\cmsss Z\kern-.4em Z}} {\lower1.2pt\hbox{\cmsss
Z\kern-.4em Z}}\else{\cmss Z\kern-.4em Z}\fi}
\def\IR{\relax{\rm I\kern-.18em R}}

\def\one{{\hbox{ 1\kern-.8mm l}}}

\def\tr{{\rm tr\,}}
\def\Tr{{\rm Tr\,}}

\newlength{\bredde}
\def\slash#1{\settowidth{\bredde}{$#1$}\ifmmode\,\raisebox{.15ex}{/}
\hspace*{-\bredde} #1\else$\,\raisebox{.15ex}{/}\hspace*{-\bredde}
#1$\fi}

\newsavebox{\zzzbar}
\sbox{\zzzbar}
  {\setlength{\unitlength}{0.9em}
  \begin{picture}(0.6,0.7)
  \thinlines
  \put(0,0){\line(1,0){0.6}}
  \put(0,0.75){\line(1,0){0.575}}
  \multiput(0,0)(0.0125,0.025){30}{\rule{0.3pt}{0.3pt}}
  \multiput(0.2,0)(0.0125,0.025){30}{\rule{0.3pt}{0.3pt}}
  \put(0,0.75){\line(0,-1){0.15}}
  \put(0.015,0.75){\line(0,-1){0.1}}
  \put(0.03,0.75){\line(0,-1){0.075}}
  \put(0.045,0.75){\line(0,-1){0.05}}
  \put(0.05,0.75){\line(0,-1){0.025}}
  \put(0.6,0){\line(0,1){0.15}}
  \put(0.585,0){\line(0,1){0.1}}
  \put(0.57,0){\line(0,1){0.075}}
  \put(0.555,0){\line(0,1){0.05}}
  \put(0.55,0){\line(0,1){0.025}}
  \end{picture}}

\def\Im{{\rm Im ~}}
\def\Re{{\rm Re ~}}

\newcommand{\ena}{\end{eqnarray}}
\newcommand{\beqa}{\begin{eqnarray}}
\newcommand{\eeqa}{\end{eqnarray}}

\def\G{\Gamma}

\def\cD{{\cal D}}

\renewcommand{\b}{\beta}
\newcommand{\g}{\gamma}

\def\d{{\rm d}}

\newcommand{\ibar}{{\bar \imath}}
\newcommand{\jbar}{{\bar \jmath}}
\newcommand{\thbar}{{\bar \theta}}
\newcommand{\Dbar}{{\bar D}}
\newcommand{\fbar}{{\bar \phi}}
\newcommand{\labar}{{\bar \lambda}}

\newcommand{\abar}{{\bar \a}}
\newcommand{\bbar}{{\bar \b}}
\newcommand{\kbar}{{\bar k}}

\newcommand{\mbar}{{\bar m}}
\newcommand{\nbar}{{\bar n}}
\newcommand{\pbar}{{\bar p}}

\newfont{\goth}{ygoth.tfm scaled 1200}                   

\def\a{\alpha}
\def\b{\beta}
\def\e{\epsilon}
\def\th{\theta}
\def\f{\phi}
\def\g{\gamma}

\def\j{\psi}

\def\m{\mu}
\def\n{\nu}

\def\s{\sigma}

\def\D{\Delta}
\def\F{\Phi}
\def\G{\Gamma}

\def\L{\mathcal{L}}
\def\O{\Omega}
\renewcommand{\O}{{\mathcal{O}}}

 \numberwithin{equation}{section}

\def\1{{(1)}}
\def\2{{(2)}}
\def\3{{(3)}}

\def\1{{\bf 1}}
\def\a{{\alpha}}

\def\M{{\mathcal M}}
\def\hetdil{\varphi_{het}}

\def\vF{{ \hat \Phi}}
\def\vP{{\hat P}}

\begin{document}
\begin{titlepage}

\begin{center}

{July 18, 2011}
\hfill         \phantom{xxx}  EFI-11-19

\vskip 2 cm {\Large \bf Linear Sigma Models with Torsion }
\vskip 1.25 cm {\bf  Callum Quigley\footnote{cquigley@uchicago.edu} and Savdeep Sethi\footnote{sethi@uchicago.edu}}\non\\
\vskip 0.2 cm
 {\it Enrico Fermi Institute, University of Chicago, Chicago, IL 60637, USA}\non\\ \vskip 0.2cm

\end{center}
\vskip 2 cm

\begin{abstract}
\baselineskip=18pt

Gauged linear sigma models with $(0,2)$ supersymmetry allow a larger choice of couplings than models with $(2,2)$ supersymmetry. We use this freedom to find a fully linear construction of torsional heterotic compactifications, including models with branes. As a non-compact example, we describe a family of metrics which correspond to deformations of the heterotic conifold by turning on $H$-flux. We then describe compact models which are gauge-invariant only at the quantum level.
Our construction gives a generalization of symplectic reduction. The resulting spaces are non-K\"ahler analogues of familiar toric spaces like complex projective space. Perturbatively conformal models can be constructed by considering intersections. 

\end{abstract}

\end{titlepage}


\section{Introduction}
\label{intro}

The most interesting class of supersymmetric string vacua are flux compactifications. Among the various ways of building four-dimensional $N=1$ string vacua, the most promising candidates for a perturbative string description are heterotic compactifications with torsion or NS three-form flux. At the level of supergravity~\cite{Strominger:1986uh, Hull:1986kz}, a torsional background requires a choice of complex manifold $\M$ with a Hermitian metric $g$ defining a fundamental form $J$,
\be
J_{m\bar{n}} = i g_{m \bar{n}},
\ee
together with a choice of holomorphic gauge bundle. For supersymmetric backgrounds, the fundamental form determines the torsion via
\be
H = i (\partial - \bar\partial) J.
\ee
Compact solutions with non-trivial $H$ are impossible at the level of supergravity. What makes such compactifications possible are $\ap$ corrections to the equations of motion and to the Bianchi identity for $H$, given by
\be
\label{bianchi} dH
= { \alpha' \over 4}  \left\{
 \tr [R(\omega_+) \wedge R(\omega_+)] - \tr [F \wedge F] \right\} + [B] \ee
where $F$ is the field strength for the gauge-bundle and $R$ is the curvature two-form, while $[B]$ denotes the cohomology class of any (anti-)NS5-brane sources. The connection used to evaluate the curvature two-forms is a combination of the usual spin connection $\omega$ and $H$:
\be \omega_+ = \omega + {1\over 2} H. \ee
The curvature correction to~\C{bianchi}\  provides a tadpole for NS5-brane charge in the heterotic background allowing for a violation of the supergravity Gauss Law constraint. It plays a role analogous to higher derivative corrections in F-theory which produce  a D3-brane tadpole, or equivalently, the role played by curvature couplings on orientifold planes in type IIB string theory.  

In the special case of a K\"ahler space, where $dJ=0$, the manifold $\M$ is Calabi-Yau at leading order in the $\ap$ expansion. This is the most heavily studied class of string compactifications. The particle physics that emerges from these spaces is quite appealing, except for the moduli problem and the issue of the cosmological constant.

Torsional backgrounds, however, are expected to have far fewer moduli. In principle, the only modulus always present is the string dilaton. The reason for this expectation is the way in which the first compact torsional metrics were found by~\cite{Dasgupta:1999ss}. Those spaces were constructed by dualizing type IIB flux vacua built on $K3\times T^2$. The IIB flux freezes out many of the geometric moduli~\cite{Dasgupta:1999ss}, and the spectrum is expected to remain unchanged after duality. The resulting torsional metrics describe a torus bundle over $K3$ twisted in a way that ensures non-K\"ahlerity. That these spaces do not admit K\"ahler metrics was nicely demonstrated in~\cite{Goldstein:2002pg}. These string backgrounds have been quite heavily studied in recent years from both space-time and world-sheet approaches, and generalized in some ways; for example, by dualizing elliptic Calabi-Yau spaces rather than $K3\times T^2$~\cite{Becker:2009df}. For a sampling of literature, see~\cite{
Becker:2002sx, LopesCardoso:2002hd,   Becker:2003yv, Gauntlett:2003cy,  Becker:2003sh,   Becker:2005nb,  Fu:2006vj, Becker:2006et, Dasgupta:2006yd, Kimura:2006af, Knauf:2006mz, Kim:2006qs, Sethi:2007bw, Becker:2007ea, Cvetic:2007ju,
Fu:2008ga,  Evslin:2008zm, Andriot:2009fp, Adams:2009zg, Adams:2009tt, Adams:2009av, Louis:2009dq, Becker:2009zx, Carlevaro:2009jx, McOrist:2010jw, Melnikov:2010pq, Gutowski:2010jk, Chen:2010bn,   Andreas:2010cv, Andriot:2011iw}.

However, our suspicion has always been that these are very special examples, much like elliptic Calabi-Yau spaces among all Calabi-Yau spaces. What we really would like is a tool to build generic torsional spaces more akin to the quintic Calabi-Yau.
The goal of this paper is to provide such a construction. A particularly useful tool for studying non-linear sigma models is the linear sigma model introduced by Witten~\cite{Witten:1993yc}. Quantities that are controlled under RG flow can be reliably computed in the ultraviolet linear theory. Among many results, the linear theory provided a tool with which the correspondence between Calabi-Yau spaces and Landau-Ginzburg theories, proposed in~\cite{Greene:1988ut}, could be studied in a concrete way.

We will generalize the linear sigma model to include torsion. It is rather crucial that we consider $(0,2)$ world-sheet theories rather than $(2,2)$ theories. This is actually the most interesting setting for studying generalized geometry because no compact torsional models exist with $(2,2)$ supersymmetry. What $(0,2)$ supersymmetry gives us is a gauge anomaly that can be used to cancel a classical violation of gauge invariance. This is central to the construction of compact torsional spaces. In terms of past work, we found particularly useful work on proving $(2,2)$ mirror symmetry~\cite{Morrison:1995yh}, and on constructing linear models for the N=2 DRS torsional backgrounds, in which the role of the gauge anomaly was explained~\cite{Adams:2006kb}.

\subsection{The basic idea and outline}

Let us draw an analogy with familiar facts from four-dimensional $N=1$ gauge theory; see, for example,~\cite{Intriligator:1995au}. The topological $\theta$-angle coupling, $\Tr( F\wedge F)$, is paired with the gauge coupling in the combination
\be
\tau = {4\pi i \over g^2} + { \theta \over 2 \pi}.
\ee
The gauge kinetic terms take the form
\be
{1\over 16 \pi}{\rm Im} \left\{  \int  d^2\theta \, \tau \, \Tr \left( W_\a W^\a \right) \right\} = - {1\over 4 g^2}  \Tr (F_{\mu\nu})^2 + {\theta \over 32 \pi^2} \Tr (F\wedge F) + \ldots.
\ee
The quantum renormalization of $\tau$ is highly constrained. Expressed in terms of a complexified strong coupling scale $\Lambda = |\Lambda| e^{i\theta/b}$, $\tau$ takes the schematic form
\be
\tau(\mu) = {b\over 2\pi i} \log\left({\Lambda\over \mu}\right) + f(\Lambda^b, \Phi),
\ee
where $b$ is determined by the one-loop beta-function, and the function $f$ is a single-valued function of chiral fields, collectively denoted $\Phi$. This form for $\tau$ respects holomorphy, the symmetry $\Lambda^b \rightarrow e^{2\pi i} \Lambda^b$ with $\tau \rightarrow \tau+1$. Note that we must introduce a scale to define the logarithm in four dimensions. Usually the logarithm is generated by integrating out physics at a higher scale.

In two-dimensional $(0,2)$ theories, reviewed in section~\ref{conventions}, there is an analogous superpotential structure where $W^\a W_\a$ is replaced by a fermionic field strength $\Upsilon$. For simplicity, let us restrict to abelian gauge theory and consider the coupling
\be
-{i \over 4} \int d\th^+ f(\Phi) \Upsilon + {\rm c.c.} =   {\rm Re} (f) D + {\rm Im}(f) F_{01} + \ldots,
\ee
where $D$ is the $D$-term auxiliary field and $F_{01}$ is the field strength. The natural periodic $\theta$-angle, given by $ {\rm Im}(f)$, is now paired with the $D$-term which determines the vacuum structure.

How shall we constrain $f$? The most straight forward case is a gauge invariant function of chirals $\Phi$. Models of this type are always non-compact and discussed in section~\ref{gaugeinvariant}. However, we could also allow a logarithm and consider
\be\label{generalf}
f(\Phi) = \sum_i N_i \log(\Phi^i) + f_0(\Phi),
\ee
where $N_i$ are integers and $f_0$ is single-valued. Unlike four dimensions, we do not need to introduce a scale to define the logarithm since two-dimensional scalar fields are dimensionless.

Such a log coupling appears problematic in the fundamental theory for two reasons: first, the theory is no longer gauge invariant if any $\Phi^i$ are charged. However, the violation of gauge invariance involves a shift proportional to $\Upsilon$, which is precisely of the type that can be canceled by a one-loop gauge anomaly. The one-loop gauge anomaly corresponds roughly to $O(\ap)$ terms on the right hand side of~\C{bianchi}\ and controls the total tadpole of the theory. The classical violation of gauge invariance from the log coupling corresponds to the flux appearing on the left hand side of~\C{bianchi}.

The second issue is defining the log at the quantum level. This looks problematic if the moduli space of the theory can access loci where singularities occur.  Fortunately, the $D$-term constraints are now also modified. Consider a model where the fields $\Phi^i$ have charges $Q^i_a$ under each $U(1)$ gauge factor labeled by $a$. The usual symplectic reduction involves solving $D$-term constraints
\be
\sum_i Q_i^a |\phi^i|^2 = r^a,
\ee
where the $r^a$ are Fayet-Iliopoulos parameters, and then quotienting by the abelian symmetry group. The log modifies these constraints as follows:
\be\label{manydterms}
\sum_i Q_i^a |\phi^i|^2 + N_i^a \log |\phi^i| = r^a.
\ee
For suitable choices of $Q_i^a$ and $N_i^a$, the singular locus of the log can be removed. This is a generalization of  symplectic reduction. After quotienting by the abelian group action, the resulting space is expected to be complex and non-K\"ahler. 

Further, we expect the theory to be well behaved on expanding around a vacuum solution to these $D$-term constraints since the log is bounded. Even when solutions of~\C{manydterms}\ permit access to singularities of the log terms, there should be an interesting interpretation in terms of throats from (anti-)NS5-brane sources, denoted $[B]$ in~\C{bianchi}. However, we suspect string perturbation breaks down near these singularities. Models of this flavor are discussed in section~\ref{noncptex}.

It is important to note that some of the $r$ parameters appearing in~\C{manydterms}\ do ${\it not}$ correspond to physical moduli. If the log interactions drop out of linear sums of the $D$-terms, we expect those combinations of $r$ parameters to correspond to moduli for conformal models. Otherwise, the physics should depend on whether the $r$ parameters lie in some range, but not on the specific value in that range. 

In section~\ref{compact}, we describe this construction in more detail and discuss issues like turning on a superpotential and the phase structure. We expect conformal models to flow to analogues of Calabi-Yau spaces, but the resulting spaces are not Ricci-flat\footnote{Even conventional $(2,2)$ sigma models do not flow to the Ricci flat metric but to a metric that differs from Ricci flat by terms higher order in $\ap$.} and instead satisfy
\bea
R_{mn}+2 \nabla_m \nabla_n
\hetdil -{1\over 4} {H}_{mpq} {{H}_n}^{pq}
-{\alpha'\over 4} \Big[\tr F_{mp}{F_n}^p   -R_{mpqr}(\omega_+)
R_{n}^{~~pqr}(\omega_+)\Big] = 0,
\eea
up to terms of order $O(\ap^2)$ when an $\ap$-expansion is valid. We expect these spaces to be topologically distinct from Calabi-Yau spaces as was the case for the metrics found in~\cite{Dasgupta:1999ss}. Here $\hetdil$ is the heterotic dilaton which is generically varying in torsional backgrounds and can give rise to large warping of the Einstein frame metric. The dilaton itself is determined by the metric and flux to ensure conformal invariance. As we discuss in section~\ref{dilatondiscussion}, there are some non-compact models where the string coupling $e^{\hetdil}$ is bounded, and some where it grows much like in the usual NS5-brane conformal field theory. 
 
Our construction also gives a natural class of supersymmetric gauge bundles over non-K\"ahler manifolds which we will not explore in detail here. Clearly, there are many interesting questions to study.
Based on intuition from type II flux vacua, it does seem likely that this class of string vacua will be significantly larger than the currently known heterotic string compactifications. 

\vskip 0.5 cm\noindent {\bf Note Added:} This work was presented at the ``Topological Heterotic Strings and (0,2) Mirror Symmetry Workshop." During that workshop, we learned about interesting independent work with related observations~\cite{Blaszczyk:2011ib}.

\section{The Basics of $(0,2)$ Models} \label{conventions}

\subsection{Chiral and Fermi superfields}

We begin by establishing our notation and conventions. For a nice review of this topic, see~\cite{McOrist:2010ae}. Throughout our discussion, we will use the language of $(0,2)$ superspace  with coordinates $(x^+,x^-,\th^+,\thbar^+)$. The world-sheet coordinates are defined by $x^\pm = \hlf(x^0\pm x^1)$, so  the corresponding derivatives $\del_\pm = \del_0 \pm \del_1$ satisfy $\del_\pm x^\pm =1$.
We define the measure for Grassmann integration so that $\d^2\th^+ = \d\thbar^+ \d\th^+$ and  $ \int\d^2\th^+\, \th^+\thbar^+ = 1.$ The $(0,2)$ super-derivatives
\be
D_+ = \del_{\th^+} - i\thbar^+\del_+, \qquad \Dbar_+=-\del_{\thbar^+} +i\th^+\del_+,
\ee
satisfy the usual anti-commutation relations
\be
\{D_+,D_+\} = \{\Dbar_+,\Dbar_+\} = 0, \qquad \{\Dbar_+,D_+\} = 2i\del_+ .
\ee

In the absence of gauge fields, $(0,2)$ sigma models involve two sets of superfields: chiral superfields annihilated by the $\Dbar_+$ operator,
\be
\Dbar_+\F^i=0,
\ee
and Fermi superfields $\G^\a$ which satisfy,
\be
\Dbar_+\G^\a = \sqrt{2} E^\a,
\ee
where $E^\a$ is chiral: $\Dbar_+ E^\a=0$. These superfields have the following component expansions:
\bea
\F^i &=& \f^i+\sqrt{2}\th^+\j_+^i - i \th^+\thbar^+\del_+\f^i, \label{chiral}\\
\G^\a &=& \g_-^\a +\sqrt{2}\th^+F^\a - \sqrt{2} \thbar^+ E^\a - i\th^+\thbar^+\del_+\g_-^\a.
\eea

If we omit superpotential couplings, the most general  Lorentz invariant $(0,2)$ supersymmetric action involving only chiral and Fermi superfields and their complex conjugates takes the form,
\be \label{(0,2) sigma}
\L =-\hlf\int\d^2\th^+\left[{i\over2}K_i \del_- \F^i - {i\over2} K_{\ibar } \del_- {\bar \F}^{\ibar} +h_{\a\bar{\b}}\bar{\G}^{\bar \b}\G^\a + h_{\a\b}\G^\a\G^\b +h_{\bar{\a}\bar{\b}}\bar{\G}^{\bar{\a}}\bar{\G}^{\bar{\b}} \right].
\ee
The one-forms $K_i$ determine the metric; they are  $(0,2)$ analogues of the K\"ahler potential which defines the simplest $(2,2)$ non-linear sigma models.  The functions $h_{\a\b}$ and $h_{\a\bar{\b}}$ determine the bundle metric.

We will not require the $E^\a$ degree of freedom for the moment so let us set $E^\a=0$. The $E^\a$ couplings introduce potential and Yukawa couplings much like a superpotential which we have also omitted. Performing the superspace integral in~\C{(0,2) sigma}\ gives the component action:
\bea\label{componentaction}
\L&=&-g_{i\jbar}\,\del_\m\f^i\del^\m\f^\jbar + b_{i\jbar}\,\e^{\m\n}\del_\m\f^i\del_\n\f^\jbar + i g_{i\jbar}\,\psi^\jbar_+\Big[\dd^i_k \del_- + \G^i_{jk}\del_-\f^j + H^i{}_{\bar{\ell} k}\del_-\f^{\bar{\ell}}\Big]\j_+^k \ \non\\
&&+ih_{\a\bar{\b}}\g_-^\bbar\Big[\dd^\a_\e\del_+ +\big(A_i\big)^\a_\e \del_+\f^i\big]\g_-^\e - {i\over2} h_{\a\bbar}\big(A_\jbar\big)^\bbar_\b\del_+\f^\jbar\g^\a_- \g_-^\b -{i\over2} h_{\a\bbar}\big(A_i\big)^\a_\bbar\del_+\f^i\g^\abar_- \g_-^\bbar \non\\
&&+\big(\cF_{i\jbar}\big)_{\a\bbar}\psi^\jbar_+\j^i_+\g^\bbar_-\g^\a_- + \hlf\big(\cF_{i\jbar}\big)_{\a\b}\psi^\jbar_+\j^i_+\g^\a_-\g^\b_- + \hlf\big(\cF_{i\jbar}\big)_{\abar\bbar}\psi^\jbar_+\j^i_+\g^\abar_-\bar{\g}^\bbar_-  \cr
&&+ h_{\a\bbar}\Big(F^\a + \big(A_i\big)^\a_\b\psi^i_+\g_-^\a + \big(A_i\big)^\a_\bbar\psi^i_+\g^\bbar_-\Big)\Big(\bar{F}^\bbar - \big(A_\jbar\big)^\bbar_{\bar{\e}}\psi^\jbar_+\g_-^{\bar{\e}} - \big(A_\jbar\big)^\bbar_\a\psi^\jbar_+\g^\a_-\Big).
\eea
The couplings appearing in~\C{componentaction}\ are given by
\bea
& g_{i\jbar} = \del_{(\jbar} K_{i)}, \qquad & b_{i\jbar} = \del_{[\jbar} K_{i]}, \cr
&\G^i_{jk} = g^{i\jbar}\del_j g_{k\jbar}, \qquad  &  H_{\ibar\jbar k} = \del_{k[\jbar}K_{\ibar]}, \cr
&\big(A_i\big)^\a_\b = h^{\a\bar{\e}}\del_i h_{\b\bar{\e}}, \qquad & \big(A_\jbar\big)^\abar_\bbar = h^{\e\abar}\del_\jbar h_{\e\bbar}, \cr
&\big(A_i\big)^\a_\bbar = -2h^{\a\bar{\e}}\del_i h_{\bbar\bar{\e}}, \qquad  & \big(A_\jbar\big)^\abar_\b = 2h^{\e\abar}\del_\jbar h_{\e \b}, \eea
and
\bea
 & \big(\cF_{i\jbar}\big)_{\a\bbar} = h_{\a\abar} \Big(\del_i\big(A_\jbar\big)^\abar_\bbar - \big(A_\jbar\big)^\abar_\b \big(A_i\big)^\b_\bbar \Big), \cr
 &\big(\cF_{i\jbar}\big)_{\a\b} = h_{\a\abar} \Big(\del_i\big(A_\jbar\big)^\abar_\b - \big(A_\jbar\big)^\abar_\e \big(A_i\big)^\e_\b \Big).
\eea
Note that the metric is in general not K\"ahler but it is always Hermitian. K\"ahlerity requires $\del_{[j}g_{k]\jbar}=0$. Note that $\G^i_{jk}$ is the Hermitian connection on the holomorphic tangent bundle and not the Levi-Civita connection. For K\"ahler manifolds,  $\G$ is symmetric in its lower indices and these two connections coincide. 







\subsection{Gauged linear sigma models}

We now need to introduce gauge fields. For a general $U(1)^n$ abelian gauge theory, we require a pair $(0,2)$  gauge superfields $A^a$ and $V_-^a$ for each abelian factor, $a=1,\ldots,n$. Let us restrict to $n=1$ for now. Under a super-gauge transformation, the vector superfields transform as follows,
\bea
\dd A &=& {i}(\bar{\La} - \La)/2, \\
\dd V_- &=& - \del_-(\La + \bar{\La})/2,
\eea
where the gauge parameter $\La$ is a chiral superfield: $\Dbar_+ \La=0$.  In Wess-Zumino gauge, the gauge superfields take the form
\bea
A &=& \th^+\thbar^+ A_+, \\
V_- &=& A_- - 2i\th^+\labar_- -2i\thbar^+\la_- + 2\th^+\thbar^+ D,
\eea
where $A_\pm = A_0 \pm A_1$ are the components of the gauge field. We will denote the gauge covariant derivatives by
\be
\cD_\pm = \del_\pm + i Q A_\pm
\ee
when acting on a field of charge $Q$. This allows us to replace our usual superderivatives $D_+,\Dbar_+$ with gauge covariant ones
\be
\mathfrak{D}_+ = \del_{\th^+} - i\thbar^+\cD_+ \qquad \bar{\mathfrak{D}}_+=-\del_{\thbar^+} +i\th^+\cD_+
\ee
which now satisfy the modified algebra
\be
\{\mathfrak{D}_+,\mathfrak{D}_+\} = \{\bar{\mathfrak{D}}_+,\bar{\mathfrak{D}}_+\} = 0 \qquad \{\bar{\mathfrak{D}}_+,\mathfrak{D}_+\} = 2i\cD_+ .
\ee
We must also introduce the supersymmetric gauge covariant derivative,
\be
\nabla_- = \del_- + i Q V_-,
\ee
which contains $\cD_-$ as its lowest component. The gauge invariant Fermi multiplet containing the field strength is defined as follows,
\be
\Upsilon =[\bar{\mathfrak{D}}_+,\nabla_-] =  \Dbar_+(\del_- A + i V_-) = -2\big(\la_- - i\th^+(D-iF_{01}) - i\th^+\thbar^+\del_+\la_-\big).
\ee
Kinetic terms for the gauge field are given by
\be \label{LU}
\L = -{1\over8e^2}\int\d^2\th^+\, \bar{\Upsilon}\Upsilon = {1\over e^2}\left(\hlf F_{01}^2 + i\labar_-\del_+\la_- + \hlf D^2\right).
\ee
Since we are considering abelian gauge groups, we can also introduce an FI term with complex coefficient $t=ir + {\th\over2\pi}$:
\be \label{LFI}
{t\over4}\int\d\th^+ \Upsilon\Big|_{\thbar^+=0} + c.c. = -rD + {\th\over2\pi}F_{01}.
\ee

In order to charge our chiral fields under the gauge action, we should ensure that they satisfy the covariant chiral constraint $\mathfrak{\bar{D}}_+\Phi = 0$. Since $\mathfrak{\bar{D}}_+ = e^{QA}\Dbar_+e^{-QA}$ it follows that $e^{QA}\Phi_0$ is a chiral field of charge $Q$, where $\Phi_0$ is the neutral chiral field appearing in \C{chiral}. In components,
\be
\F = \f + \sqrt{2} \th^+ \j -i\th^+\thbar^+\cD_+\f
\ee
The standard kinetic terms for charged chirals in $(0,2)$ gauged linear sigma models (GLSMs) are
\bea \label{LPhi}
\L &=& {-i\over2}\int\d^2\th^+\ \bar{\F}^i \nabla_- \F^i, \\
&=& \left(-\big|\cD_\mu \f^i\big|^2 + \bar{\psi}_+i\cD_-\psi_+^i - \sqrt{2}iQ_i \bar{\f}^i\la_-\j^i_+ + \sqrt{2}iQ_i\f^i\bar{\j}_+^i\labar_- + Q_i \big|\f^i\big|^2\right). \non
\eea
Fermi superfields are treated similarly. We promote them to charged fields by defining $\G = e^{QA}\G_{0}$ so that in components
\be
\G = \g_- + \sqrt{2}\th^+F - \sqrt{2}\thbar^+E -i\th^+\thbar^+\cD_+\g_-,
\ee
where we have introduced a non-vanishing $E$ again.
If we make the standard assumption that $E$ is a holomorphic function of the $\F^i$, then the standard kinetic terms for the Fermi fields are
\bea \label{LLa}
\L &=& -\hlf\int\d^2\th^+\,  \bar{\G}^\a \G^\a, \\
&=& \left(i\bar{\g}_-^\a\cD_+\g_-^\a + \big|F^\a\big|^2 - \big|E^\a\big|^2 - \bar{\g}^\a_-\del_i E^\a \j_+^i - \bar{\j}_+^i \del_\ibar \bar{E}^\a \g^\a_-\right). \non
\eea
It is also possible to add a superpotential to the theory, but we will postpone adding that coupling until a later section. In the absence of any superpotential couplings, the action consisting of the terms \C{LU}, \C{LFI}, \C{LPhi} and \C{LLa} comprises the standard $(0,2)$ GLSM.

\subsection{The classical IR geometry}
\label{classicalgeometry}
The classical infra-red limit of a  $U(1)^n$ GLSM corresponds to sending $e_a\rightarrow\infty$, since these gauge couplings are  dimensionful quantities. In this limit, formally the $\Upsilon^a$ kinetic terms disappear resulting in the simple on-shell bosonic action,
\be
\L_B = - \big|\del_\mu\f^i\big|^2 + j^{a\m}A_\m^a -\hlf \left(\D^{-1}\right)^{ab} A^a_\m A^{b\m} +{\th^a\over2\pi}F_{01}^a - V(\phi), \label{bosonic}
\ee
where
\bea
j^a_\mu &=& i\sum_i Q^a_i (\fbar^i\del_\mu \phi^i - \f^i\del_\mu\fbar^i), \\
\left(\D^{-1}\right)^{ab} &=& 2\sum_i Q_i^a Q_i^b |\f^i|^2,
\eea
and the scalar potential is
\be
V(\f) = |E^\a(\f)|^2 + \sum_a {1\over 2e_a^2} D^a D^a
\ee
with
\be
D^a= -{e_a^2} \Big(\sum_i Q_i^a |\f^i|^2  - r^a \Big).\label{D}
\ee
Let us once again consider the case where all $E^\a$ are zero and assume $N$ fields $\phi^i$. The vacuum manifold of the theory is then the toric variety $X = D_a^{-1}(0)/U(1)^n$. That is, $X$ is the $N-n$ dimensional space given by the symplectic quotient $\CC^N//U(1)^n$ with moment maps $D^a$.

We can extract a lot of geometric data about $X$ by considering the low-energy effective action for the GLSM, which is classically a non-linear sigma model with target $X$.
To see this, note that the gauge field becomes non-dynamical in this limit so we can solve for it algebraically,
\be
A_\mu^a = \D^{ab}{j}^b_\mu. \label{A}
\ee
Notice that under a gauge transformation $A^a\rightarrow A^a-\d\La^a$ as it should. However, rather than interpret $A^a$ as a collection of gauge connections as we do in the linear theory, we now view them as (pullbacks of) connections on a set of line bundles $L^a$ over $X$. The gauge transformations should now be viewed as defining the $L^a$ across patches. The curvature of these line bundles, $F^a$, are elements of $H^2(X,\Z)$ and it is straightforward to show that the class of the complexified K\"ahler form of $X$ is given by
\be
[\cJ] = [B] + i[J] = (\th^a + i r^a)[F^a] = t^a[F^a].
\ee
Indeed, after substituting~\C{A}\ for $A^a$ into the bosonic action~\C{bosonic}\ and making use of the $D$-term constraint~\C{D}, we find the target space metric
\be
\d s^2 = \big|\d\f^i\big|^2  - 2\D^{ab}\left(\sum_i Q_i^a\bar{\f}^i\d\f^i\right)\left(\sum_j Q_j^b\f^j\d\bar{\f}^j\right),
\ee
which generalizes the Fubini-Study metric. The pullback of $B$ from $X$ to the world-sheet is given by,
\be
B = \e^{\mu\nu}B_{i\jbar}\, \del_\mu\phi^i\del_\nu\phi^\jbar = {\th^a\over2\pi} F^a.
\ee
In this class of models $X$ is always K\"ahler. This follows directly from the fact that $J$ lies in $H^2(X,\Z)$ and so is closed.  Since $B$ lies in the same cohomology class, it is also closed: $H=\d B=0$. To find manifolds with torsion, we must generalize the symplectic quotient in a suitable manner.

\section{Non-Compact Models}
\label{gaugeinvariant}
\subsection{Gauge Invariant $f$}
As we discussed in the introduction, the simplest way to include torsion in a GLSM is to make the FI terms field-dependent. In this section, let us add the couplings
\be\label{f}
-{i\over4}\int\d^2x\d\th^+\, f^a(\F)\Upsilon_-^a + c.c.
\ee
and restrict our attention to gauge-invariant $f^a$. The case of gauge non-invariant $f^a$ needed for compact models will be considered in section~\ref{compact}.
Since the $f^a$ are required to be gauge invariant, this forces us to introduce fields with negative charges.\footnote{One might also consider rational functions which are gauge invariant. This will generically introduce singularities but it might be possible to excise the singular loci with a suitable superpotential. We will restrict to globally defined $f^a$ in this section. } This means these models will always be non-compact in the absence of superpotential couplings.

It is easy to see that including these generalized FI terms modifies the bosonic action \C{bosonic} simply by replacing
\bea
&&r^a \rightarrow R^a(\f) = r^a - \Re(f^a), \\
&&\th^a \rightarrow \Th^a(\f) = \th^a +2\pi\Im(f).
\eea
Again we solve for the gauge fields at low energies and interpret them as connections on a set of line bundles $L^a$:
\bea
\tilde{A}^a_\m = \D^{ab}\Big(j^b_\m + {1\over2\pi}\e_{\m\n}\del^\n\Th^b\Big) = A^a_{\mu} + A'^a_\mu.
\eea
We have split the connection into a term $A^a$ from $j$ which transforms under the gauge symmetry, $\dd A^a = -\d\La^a$, and a term $A'^a$ from $\Th$ which is invariant.
 While both terms contribute to the curvature of the associated bundle,
\be
\tilde{F}^a = F^a + F'^a,
\ee
only the first term is non-trivial in cohomology since $A'$ is globally defined. Thus
\be
[\tilde{F}^a] = [F^a],
\ee
and it is $F^a$ which will appear in the complexified fundamental form\footnote{This is the two-form which would be the (complexified) K\"ahler form if it were closed.}
\be
\cJ = B + i J = (\Th^a + iR^a)F^a.
\ee
It is clear that $\cJ$ is not closed so $X$ does not inherit a K\"ahler form by reduction. It would be interesting to understand whether these spaces can ever admit a K\"ahler metric when there is non-trivial torsion $H\neq0$. Since $f^a(\phi)$ is gauge invariant it follows that $\cJ$ is globally defined; hence the class of $\d\cJ$ is trivial:
\be
[\d\cJ] = [H] + i[\d J] = [d(\Th^a+iR^a)\wedge F^a] = 0.
\ee
This can be seen more explicitly by plugging the solution for $\tilde{A}^a$ back into the bosonic action and reading off the target space metric and $B$-field from the sigma model action. To get a Hermitian metric on $X$, it is necessary to use,
\be
\d r^a = \sum_i\big(Q^a_i\bar{\f}^i + \hlf f^a_i\big)\d\f^i + \sum_i\big(Q^a_i \bar{\f}^i + \hlf \bar{f}^a_\ibar\big)\d\bar{\f}^i = 0,
\ee
where $f_i = \del_i f$, in order to swap some holomorphic and anti-holomorphic differentials. We then find the metric
\be \label{herm}
\d s^2 = \big|\d\f^i\big|^2  - 2\D^{ab}\big(\sum_i Q^a_i\bar{\f}^i\d\f^i\big)\big(\sum_i Q^b_i\f^i\d\bar{\f}^i\big) + \hlf \D^{ab}\big( f^a_i\d\f^i)( \bar{f}^b_\ibar\d\bar{\f}^i )
\ee
and $B$-field
\be\label{B1}
B =  {1\over2\pi}(\D^{ab}j^a)\wedge \d\Th^b \simeq {\Th^a\over2\pi}F^a,
\ee
where we have shifted $B$ by an exact two-form to arrive at the right hand side. Note that $f^a$ is continuously tunable in these models, which gives a tunable $H$-field which is permitted in a non-compact model. 

\subsection{An alternate derivation of the sigma model couplings}

The preceding discussion of non-compact torsional models obscures many of their important properties. Finding a hermitian metric required use of the $D$-term constraint. It is also not immediately clear that the torsion satisfies $H=i(\del-\delbar)J$, which must be true for a supersymmetric background.
These are properties required by world-sheet supersymmetry so we should expect that by working with manifest $(0,2)$ susy, rather than just the bosonic terms in the action, these features will emerge naturally. Indeed this is the case, as we will now show.

Recall in section~\ref{conventions}, we showed that the (Hermitian) metric and $B$-field of any $(0,2)$ non-linear sigma model are derived from one quantity. The superspace action (for the chiral fields only)
\be \label{action}
\L = -{i\over4}\int\d^2\th^+\ \left(K_i(\F,\bar{\F})\del_-\F^i - K_\ibar(\F,\bar{\F})\del_-\F^\ibar\right)
\ee
is determined by the $(1,0)$ form $K = K_i\d\f^i$ with complex conjugate  $K^*=K_\ibar\d\f^\ibar$. The $1$-form $K$ is the $(0,2)$ analogue of the K\"ahler potential. The target space fields are determined by $K$,
\be \label{GandB}
G_{i\jbar} = K_{(i,\jbar)} \qquad and \qquad B_{i\jbar} = K_{[i,\jbar]}.
\ee
Clearly any $(0,2)$ theory for which $K=\del k$ for some scalar function $k$ is actually K\"ahler with K\"ahler potential $k$.  The  $(0,2)$ analogue of a K\"ahler transformation is
\be \label{K1}
K(\F,\bar{\F}) \rightarrow K(\F,\bar{\F}) + K'(\F)
\ee
where $ K'(\F)$ is a holomorphic $(1,0)$-form. These transformations leave the physical couplings in \C{GandB} invariant. Furthermore, shifts in $K$ of the form
\be \label{K2}
K \rightarrow K +i\,\del U,
\ee
for any real valued function $U$, shift the Lagrangian~\C{action}\ by a total derivative and so are also symmetries.

To find the $K$ governing the classical IR geometry, we again consider the $e_a\rightarrow\infty$ limit. With the $\Upsilon^a$ kinetic terms decoupled, the superspace action is just
\bea
\L &=& {-i\over4}\int\d^2\th^+\, \left(\bar{\F}^i\nabla_-\F^i  -c.c\right) -{i\over4} \left(\int\d\th^+\left(it^a+f^a(\F)\right) \Upsilon^a -c.c.\right), \\
&=& {-i\over4}\int\d^2\th^+\, \left(\bar{\F}^i_0 e^{Q^b_i A^b} \del_-\left(e^{Q^b_iA^b}\F^i_0\right) + \left(it^a +f^a(\F)\right) \del_-A^a\right)  +c.c.\non\\
&&+\, \hlf\int\d^2\th^+\, \left(\sum_i Q^a_ie^{2Q_i^b A^b}|\F^i_0|^2 +\Re(f^a)  -r^a  \right)V_-^a, \non
\eea
where we have used the relation $\Upsilon = \Dbar_+(\del_-A + i V_-) = -\int\d\thbar^+(\del_-A + i V_-)$ up to a total derivative. Now $V_-^a$ appear as Lagrange multipliers which we can integrate out to obtain the constraints
\be \label{constraint}
\sum_i Q^a_i|\F^i|^2 e^{2Q^b_iA^b}  + \Re(f^a)  = r^a,
\ee
where we have dropped the ``0" subscripts from the uncharged $\F^i$. This superfield constraint contains the solutions for both $\tilde{A}^a$ and $D^a$ from the previous section in its component expansion. The superfield $A$ can now be eliminated from the action by using~\C{constraint}\ to solve for $A=A(\F,\bar{\F})$ implicitly. The result is a non-linear sigma model for $\Phi^i$ specified by
\be
K_i = \bar{\F}^i e^{2Q^a_iA^a(\F,\bar{\F})} + {i\over2\pi} \Th^a \del_i A^a(\F,\bar{\F}).
\ee
Adding a total derivative, we can write this as
\be
K_i \simeq \bar{\F}^i e^{2Q^a_iA^a(\F,\bar{\F})}  - {i\over2\pi} A^a(\F,\bar{\F})\del_i\Th^a .
\ee
In particular,
\bea
G_{i\jbar} &=& K_{(i,\jbar)} =  \dd_{i\jbar}\, e^{2Q^a_iA^a}  +\big(Q^a_i\bar{\f}^ie^{2Q^b_iA^b} - f_i\big)\del_\jbar A^a + \big(Q^a_j\f^j e^{2Q^b_jA^b} - \bar{f}_\jbar\big)\del_i A^a  \label{G}\\
B_{i\jbar} &=& K_{[i,\jbar]} =  \big(Q^a_i\bar{\f}^ie^{2Q^b_iA^b} - f_i\big)\del_\jbar A^a  - \big(Q^a_j\bar{\f}^j e^{2Q^b_jA^b} - \bar{f}_\jbar\big)\del_i A^a. \label{B}
\eea
One advantage of this approach is that the fundamental 2-form
\be
J = {i\over2}(\delbar K - \del K^*) = iG_{i\jbar}\,\d\f^i\wedge\d\f^\jbar
\ee
is automatically related to $H$ in the desired manner,
\be
H=(\del+\delbar) B = -\hlf(\del\delbar K + \delbar\del K^*) = i(\del-\delbar)J,
\ee
so these models always manifestly preserve target space supersymmetry. The components of $H$, given by
\bea
H_{ij\bar{k}} 
= {1\over4}\bar{\f}^j\del_{i\bar{k}}e^{2Q^a_j A^a} -{1\over4}\bar{\f}^i\del_{j\bar{k}}e^{2Q^a_i A^a} -Q^a_ke^{2Q^b_kA^b}\dd_{\bar{k}[i}\del_{j]}A^a +f^a_{[i}\del_{j]\kbar}A^a \label{H},
\eea
are generally non-vanishing. Additionally, the $(3,0)$ component of $H$ is automatically zero here. To see this, we can trivialize the $(3,0)$ component of $H$ locally with respect to a $(2,0)$ $B$-field but $B^{2,0} = \del K $, and therefore $H^{3,0}=\del B^{2,0} = \del^2 K =0$.

\subsection{A special case corresponding to UV $B$-fields}

The case of quadratic $f^a$ is particularly interesting. In this case, we can rewrite the superpotential coupling~\C{f}\ as a $D$-term that preserves linearity of the theory. Since $f^a$ is quadratic, we require pairs of fields with equal and opposite charge, $\F^i$, $\F^j$ where $Q^a_i=-Q^a_j$. Notice that we can now write $f_{ij}^a = Q_j^a b_{ij}$ for some \textit{anti-symmetric} $b_{ij}.$\footnote{While $f_{ij}^a$ is symmetric in $i,j$, $b_{ij}$ must be anti-symmetric because $Q_i^a=-Q_j^a$.} We now see that
\bea
\int\d^2x\d\th^+\, (f^a_{ij} \F^i\F^j)\Upsilon^a &=& \int\d^2x\d\th^+\, (Q^a_jb_{ij} \F^i\F^j)\Upsilon^a, \non\\
&=& \int\d^2x\d\th^+\, \Dbar_+ \big(b_{ij}\F^i\nabla_-\F^j\big), \\
&=& \int\d^2x\d^2\th^+\, b_{ij}\F^i\nabla_-\F^j.  \non
\eea
Only for this case of quadratic $f^a$ can we equivalently write these generalized FI couplings as a choice of UV $B$-field coupling,
\be \label{b}
\L = {i\over4}\int\d^2\th^+\left(b_{ij} \Phi^i\nabla_-\Phi^j - b_{\ibar \jbar} \bar{\Phi}^i\nabla_-\bar{\Phi}^j \right)
\ee
with $b_{ij} = -b_{ji} = b_{\ibar\jbar}^*$. In fact,~\C{b}\ is the most general non-trivial linear deformation of $K_i$ consistent with gauge invariance.\footnote{The other possibility, $K_i = b_{i\jbar}\Phi^\jbar$,  contributes a total derivative.} We should also point out that this coupling does not appear in a $(2,2)$ theory constructed from chiral superfields. The simplest K\"ahler potential one might try, $K = b_{ij}\Phi^i\Phi^j$, vanishes by anti-symmetry. Even if one splits the fields into groups of positively charged $\Phi^i$  and negatively charged $\Phi^a$ then
\be
K = b_{ia}\Phi^i\Phi^a + c.c. = b_{ia}\Phi^i_0 e^{(Q_i+Q_a)V}\Phi^a_0 +c.c.= b_{ia}\Phi^i_0\Phi^a_0 +c.c.
\ee
can be gauged away by a K\"ahler transformation. Usually a $B$-field in closed string theory with trivial target space and a flat metric has no effect on the physics. Indeed~\C{b}\ is trivial for neutral fields since a holomorphic deformation of $K_i$ does not alter the physical couplings. Only the presence of the (real) gauge field $V_-$ makes this coupling non-holomorphic and relevant for the low-energy physics.

We suspect this form for the field-dependent FI parameters might be useful for implementing world-sheet duality along the lines of~\cite{Adams:2003zy}.  This quadratic case is also special because of the behavior of the dilaton, which we will discuss shortly. 

\subsection{An example: the conifold with torsion}
\label{conifoldsec}
\subsubsection{Quadratic $f$}

Let us use the conifold as a nice non-compact example. Take a single $U(1)$ gauge group coupled to two chiral fields $\F^i$  $(i=1,2)$ with charge $Q_i=+1$, and two fields $\F^m$ $(m=1,2)$ of charge $Q_m=-1$.\footnote{We will ignore the fermionic sector for now, though appropriately charged left-moving fermions should be included to cancel the gauge anomaly.} In the absence of any $f(\F)$ coupling, the $D$-term condition is
\be
|\f^i|^2 - |\f^m|^2  = r.
\ee
The target space of this GLSM is the total space of the vector bundle $\O(-1)\oplus\O(-1)$ over $\PP^1$. The size of the $\PP^1$ base is controlled by $r$. In the limit $r\rightarrow0$, the space develops a conifold singularity, while finite $r$ corresponds to a resolved conifold. 

Let us restrict to a quadratic $f= f_{im}\F^i\F^m$. In this example, the superfield constraint~\C{constraint}\  becomes
\be\label{Aconstraint}
e^{2A}|\F^i|^2 - e^{-2A}|\F^m|^2 + Re(f_{im}\F^i\F^m) = r.
\ee
Introduce the notation
\be
x = |\F^i|^2,\qquad y = |\F^m|^2,\qquad z = Re(f_{im}\F^i\F^m),
\ee
and
\be
\f_i = \bar{\f}^i,\qquad \f_\ibar = \f^i, \qquad \tilde{\f}_i = f_{im}\f^m, \qquad \tilde{\f}_\ibar = \bar{f}_{im}\bar{\f}^m.
\ee
We can now solve~\C{Aconstraint}\ for $A$:
\bea
e^{2A} ={r-z+\sqrt{(r-z)^2+4xy}\over2x} = {2y\over z-r+\sqrt{(z-r)^2+4xy}}.
\eea
Plugging this expression for $A$ into the formulae~\C{G},~\C{B}, and~\C{H}\ for the target space fields gives the metric
\bea
G_{i\jbar} &=& e^{2A}\dd_{i\jbar} - { e^{4A} \f_i\f_{\jbar} - \tf_i\tf_\jbar\over \sqrt{(r-z)^2+4xy}}, \non\\
G_{i\mbar} &=& {\f_i\f_\mbar - \tf_i\tf_\mbar \over \sqrt{(r-z)^2+4xy}}, \\
G_{m\nbar} &=& e^{-2A}\dd_{m\nbar} - { e^{-4A} \f_m\f_{\nbar} - \tf_m\tf_\nbar\over \sqrt{(r-z)^2+4xy}} \non,
\eea
and $B$-field
\bea
B_{i\jbar} &=& - {e^{2A}(\f_i\tf_\jbar - \f_\jbar\tf_i) \over \sqrt{(r-z)^2+4xy}}, \non\\
B_{i\mbar} &=& {e^{2A}\f_i\tf_\mbar - e^{-2A}\f_\mbar\tf_i \over \sqrt{(r-z)^2+4xy}}, \\
B_{m\nbar} &=& -{e^{-2A}(\f_m\tf_\nbar - \f_\nbar\tf_m) \over \sqrt{(r-z)^2+4xy}} \non,
\eea
with $H$-flux
\bea
H_{ij\kbar} &=& {e^{2A}\dd_{\kbar[i}\tf_{j]} \over \sqrt{(r-z)^2+4xy}} + {e^{4A}(z-r+2\sqrt{(r-z)^2+4xy})\f_{[i}\tf_{j]}\f_\kbar - (2y) \f_{[i}\tf_{j]}\tf_\kbar \over \big((r-z)^2+4xy\big)^{3\over2}}, \non\\
H_{ij\mbar} &=& {(z-r)\f_{[i}\tf_{j]}\f_\mbar + (2y) \f_{[i}\tf_{j]}\tf_{\mbar} \over \big((r-z)^2+4xy\big)^{3\over2}}, \non\\
H_{im\jbar} &=& -{e^{2A}\tf_m\dd_{i\jbar} \over 2\sqrt{(r-z)^2+4xy}} \\
&+& {e^{4A}(z-r+2\sqrt{(r-z)^2+4xy})\f_i\tf_m\f_\jbar + (r-z)\f_m\tf_i\f_\jbar + (2y)\f_i\tf_m\tf_\jbar + (2x)\f_m\tf_i\tf_\jbar \over 2\big((r-z)^2+4xy\big)^{3\over2}}, \non\\
H_{im\nbar} &=&{e^{-2A}\tf_i\dd_{m\nbar} \over 2\sqrt{(r-z)^2+4xy}} \non\\
&+& {e^{-4A}(z-r-2\sqrt{(r-z)^2+4xy})\f_m\tf_i\f_\nbar + (r-z)\f_i\tf_m\f_\nbar - (2y)\f_i\tf_m\tf_\nbar - (2x)\f_m\tf_i\tf_\nbar \over 2\big((r-z)^2+4xy\big)^{3\over2}}, \non\\
H_{mn\jbar} &=& {(r-z)\f_{[m}\tf_{n]}\f_\jbar + (2x) \f_{[m}\tf_{n]}\tf_\jbar \over 2\big((r-z)^2+4xy\big)^{3\over2}}, \non\\
H_{mn\pbar} &=& {e^{-2A}\dd_{\pbar[m}\tf_{n]} \over \sqrt{(r-z)^2+4xy}} + {e^{-4A}(z-r-2\sqrt{(r-z)^2+4xy})\f_{[m}\tf_{n]}\f_\pbar -  (2x) \f_{[m}\tf_{n]}\tf_\pbar \over \big((r-z)^2+4xy\big)^{3\over2}} \non.
\eea
The $D$-term constraint gives the relation
\be
x-y  = r-z
\ee
which  implies
\be
e^{2A}=1\qquad and \qquad\sqrt{(r-z)^2+4xy} = x+y.
\ee
Using these relations puts the metric and $B$ into the form we expect from~\C{herm}\ and~\C{B1}\ with metric 
\bea
G_{i\jbar} &=& \dd_{i\jbar} - { \f_i\f_{\jbar} - \tf_i\tf_\jbar\over \sum |\f|^2}, \non\\
G_{i\mbar} &=& {\f_i\f_\mbar - \tf_i\tf_\mbar \over \sum |\f|^2 }, \\
G_{m\nbar} &=& \dd_{m\nbar} - {  \f_m\f_{\nbar} - \tf_m\tf_\nbar\over \sum |\f|^2} \non,
\eea
and $B$-field
\bea
B_{i\jbar} &=& - {\f_i\tf_\jbar - \f_\jbar\tf_i \over \sum |\f|^2}, \non\\
B_{i\mbar} &=& {\f_i\tf_\mbar - \f_\mbar\tf_i \over\sum |\f|^2}, \\
B_{m\nbar} &=& -{\f_m\tf_\nbar - \f_\nbar\tf_m \over \sum |\f|^2} \non.
\eea

Non-K\"ahler metrics describing flux deformations of the conifold have been obtained from a space-time perspective in~\cite{Butti:2004pk, Casero:2006pt, Maldacena:2009mw, Martelli:2010jx, Carlevaro:2009jx}. It would be interesting to connect this class of world-sheet models with those solutions. In particular, it will be very interesting to see whether the metrics and $B$-fields emerging from this construction actually solve the space-time equations of motion. 

\subsubsection{General $f$ and growth of the dilaton}
\label{dilatondiscussion}

This conifold example is conformal in the absence of $f$. Since $f$ is a superpotential coupling, we do not expect any renormalization of this coupling. At least naively, any choice of gauge-invariant $f$ would seem to give a deformation that preserves (perturbative) conformal invariance. That leads to an enormous class of non-compact models smoothly connected to any non-compact toric Calabi-Yau space. It would be very surprising if all such models corresponded to perturbative string backgrounds.  

Notice that only in the case of quadratic $f\sim \f^2$ are the metric and $B$-field homogeneous in $\f$. For $f\sim \f^n$ for $n>2$, these fields along with $H$ grow unbounded as $|\f|\rightarrow\infty$.\footnote{Actually, the critical exponent for $H$ to grow at infinity is $n=5/2$ but restricting to polynomial $f$, this amounts to the same thing.} For example, in the case of our deformed conifold at large values of $|\f|$, the flux looks like
\be
H \longrightarrow {|\f|^2\f\,\d\f\, |\d f|^2 \over (\f\fbar)^6} \sim \f |\f|^{2(n-3)} (\d\f)^3.
\ee
It is easy to see that $*H \sim \sqrt{g}(g^{-1})^3H$ will have a similar behavior. However, for a heterotic string background the dilaton $\hetdil$ and $H$ are related via the equation of motion
\be \label{fixdilaton}
\d\Big(e^{-2\hetdil}*H\Big) = O(\ap). 
\ee
For this conifold example, this relation can only be satisfied if $e^{2\hetdil}\sim \f |\f|^{2(n-3)}$ for large values of $\f$. However, for $n\geq{5\over2}$ the string coupling diverges and the world-sheet theory no longer defines a perturbative string background.

\section{Including Anomalous Couplings}
\label{compact}

\subsection{The condition for anomaly cancelation}

To construct compact models, we are interested in couplings we can add to the classical action which are \textit{not}\ gauge invariant. The classical violation of gauge invariance must be of a form that matches the quantum one-loop gauge anomaly. The sign of the anomaly is rather important for us, so we have presented a derivation of the anomaly in Appendix~\ref{anomalyapp}. The anomaly shifts the action by
\be\label{minimalgauge}
\dd S = {\cA^{ab} \over {4\pi}}\int\d^2x  \La^aF^b
\ee
where $\Lambda^a$ is the gauge parameter,  and
\be
\cA^{ab} = \sum_i Q_i^a Q_i^b - \sum_\a Q_\a^a Q_\b^b
\ee
is the anomaly coefficient with  charges $Q_i$ for right-movers and  charges $Q_\a$ for left-movers. In superspace, this reads
\be
\dd S = \left( {\cA^{ab} \over {16\pi}}\int\d^2x \d\th^+\, \La^a \Upsilon^b + c.c. \right). 
\ee
Note that a background NS5-brane can be viewed as a small instanton in the gauge bundle and so would shift the action like a left-mover. An anti-NS5-brane would  induce a shift with opposite sign. The sign of the anomaly  determines whether a positive or negative coefficient of the log corresponds to NS5-brane or anti-NS5-brane flux which is why the sign is of importance for us. 

In conventional $(0,2)$ models, the overall sign of the anomaly is unimportant since we just need to ensure the quantum anomaly vanishes. In our case, we are canceling the non-gauge invariance from the log pre-factor of $\Upsilon$ appearing in~\C{generalf}\ against both classical couplings described below and the anomaly. 

There are basically two classical couplings that we can consider. The first is the log-type FI coupling
\be\label{logfi}
S_1 = -{i \over 8\pi}\int\d^2x \d\th^+\, N_i^{a} \log\left( \F^i \right)   \Upsilon^a + c.c.
\ee
for some choice of  $N_i^{a}$. The simplest assumption is to take $N^i_a\in \Z$. This ensures invariance under the global transformation $\Phi^i \rightarrow e^{2\pi i} \Phi^i$ in any topologically non-trivial instanton sector. However, this appears to be too strong a condition. To cancel the basic the minimal gauge anomaly for a charge one left or right-mover given in~\C{minimalgauge}, we actually need to allow half-integer $N^i_a$. 

How this weaker condition is consistent in odd charge instanton sectors is a fascinating question; we will not pursue this question here, beyond commenting that perhaps an odd number of fermion zero modes in those sectors kills the path-integral rendering the theory consistent.   It will also be very interesting to see if the instanton analysis leading to the usual quantization condition on the $N^i_a$ is modified by the dynamical theta angles which, in turn, could relax the half-integrality condition further.
 We will see that the quantization of $N_i^a$ leads to a quantized $H$-flux unlike the models of section~\ref{gaugeinvariant}. Under a gauge transformation, this term will shift the action by the following amount
\be
\dd S_1 = \left( {N_{i}^{a} Q_i^b \over {8\pi}}\int\d^2x \d\th^+\, \La^b \Upsilon^a + {c.c.} \right).
\ee
Notice that only the symmetric part of $N_i^{a} Q_i^b$ can be canceled by the anomaly since $\cA^{ab}$ is manifestly symmetric. 

One might imagine replacing the monomial argument of the log in~\C{logfi}\ with a more complicated function with definite charge under the gauge symmetries like a polynomial. The difficulty with such a choice is ensuring invariance of the theory under $\Phi^i \rightarrow e^{2\pi i} \Phi^i$ for each $i$ separately. It would be very interesting if cases generalizing the monomial (or product of monomials) could be made sensible.

To produce an antisymmetric shift, consider the following term
\be\label{s2}
S_2 = {1\over4\pi} \int\d^2x\d^2\th^+\, T^{ab} A^a V_-^b
\ee
where $T^{ab}$ is an antisymmetric tensor to be determined. The $(2,2)$ extension of this coupling interestingly appeared in~\cite{Morrison:1995yh}. Under a gauge transformation,
\bea
\dd S_2 &=& {1\over4\pi} T^{ab}\int\d^2x\d^2\th^+\left({i\over2}(\bar{\La}^a - \La^a)V_-^b - \hlf A^a\del_-(\La^b+\bar{\La}^b) + {i\over4}\left(\La^a-\bar{\La}^a)\del_-(\La^b+\bar{\La}^b) \right)\right) \non\\
&=& -{1\over4\pi} T^{ab}\int\d^2x\d^2\th^+\left(\hlf\La^a(\del_-A^b + iV_-^b) + \hlf \bar{\La}^a (\del_-A^b - iV_-^b) \right) \non \\
&=& \left(- {1\over8\pi} T^{ab}\int\d^2x\d\th^+\, \La^a\Upsilon^b + c.c. \right).
\eea
Note that the terms quadratic in $\La^a$ either cancel after integration by parts or are purely (anti)-holomorphic and so only contribute a total derivative. Comparing $\dd S_1$ and $\dd S_2$ we see that $T$ must be chosen so that
\be
T^{ab} = Q_i^{[a} N_{i}^{b]}.
\ee
Together the classically anomalous terms in the action take the form
\be
S_{anom} = {1\over4\pi}\int\d^2x\left[\d^2\th^+ Q_i^{[a} N_{i}^{b]} A^a V_-^b - \left({i\over2} N_i^{a} \int\d\th^+\, \log (\F^i) \Upsilon^a +c.c.\right)\right].
\ee
Under a gauge transformation,
\bea
\dd S_{anom} &=& {Q_i^aN_i^{b}-Q_i^{[a} N_{i}^{b]}\over8\pi}\int\d^2x\d\th^+\, \La^a\Upsilon^b + c.c. \\
&=&{Q_i^{(a}N_{i}^{b)}\over8\pi}\int\d^2x\d\th^+\, \La^a\Upsilon^b + c.c. \non,
\eea
so the requirement of a consistent theory is
\be
\sum_i Q_i^{(a}N_{ i}^{b)} +{1\over 2}\cA^{ab}=0.
\ee
So far, our discussion is largely focused on the classical physics of these models along with the quantum condition for gauge invariance. Standard $(0,2)$ theories are perturbatively conformal if the $\sum_i Q_i^a=0$ for each $a$. Since we are modifying a superpotential coupling, albeit with a log, we suspect that this condition is unchanged as long as the theory has a moduli space that excludes singularities of the log couplings. We will see later that there are many choices of $N^{a}_i$ for which this is the case.    

If one is uncomfortable with the log interaction, it can be replaced by more familiar couplings as follows:\footnote{We would like to thank Allan Adams for suggesting this replacement.} for each $\F^i$, introduce an axially gauged field $Y^i$ transforming in the following way under a gauge transformation
\be
Y^i \rightarrow Y^i + i Q_i^a \La^a.
\ee
Now consider the couplings 
\be \label{defY}
S_Y = -{i \over 8\pi}\int\d^2x \d\th^+\,  \left( N_i^{a} Y^i   \Upsilon^a +  \G_{Y^i} \left\{ e^{Y^i} - \Phi^i \right\} \right) +  c.c.
\ee
where $\G_{Y^i}$ are standard chiral Fermi superfields. Solving the superpotential constraint from $\G_{Y^i}$ sets $e^{Y^i} = \Phi^i$. This form again suggests that the renormalization of the theory should not be problematic as long as singular loci are excluded from the moduli space. It is worth noting that the metric expressed in terms of $Y$-fields is not flat. One could also consider a flat metric for the $Y$-fields which leads to models of the type studied in~\cite{Adams:2006kb}.  

\subsection{Supersymmetry anomaly}
\label{susyanomaly}

Introducing log interactions that break gauge invariance also leads to a classical breaking of $(0,2)$ supersymmetry. This is surprising since the action expressed in superspace appears supersymmetric. Indeed the theory is supersymmetric if we choose not to fix Wess-Zumino gauge and consider a theory with extra degrees of freedom in the vector multiplet which would usually decouple with this gauge choice. However, choosing Wess-Zumino gauge is not compatible with preserving supersymmetry. Rather a compensating gauge transformation must accompany a supersymmetry transformation in order to preserve this gauge choice. This is the basic source of the supersymmetry anomaly. It is tied directly to the gauge anomaly. 

In terms of standard physical fields, we can see this directly from the action as follows: imagine a single charged scalar $\Phi$ with charge $Q$ and the superpotential coupling
\be\label{susyissue}
\int\d\th^+\, \log (\F) \Upsilon  = 2i (D-iF) \log (\f) - 2\sqrt{2} {\psi_+ \lambda_- \over \phi} .
\ee
The problematic non-cancelation comes from the variation
\be
\delta \psi_+ = \sqrt{2} i  {\bar \e} \, \cD_+\f.
\ee
If $\phi$ were neutral then $\cD_+ \rightarrow \p_+$ and the variation of the second term in~\C{susyissue}\ would cancel against the variation of the first term up to a total derivative.  This is no longer the case when $\phi$ is charged and we pick up a term proportional to $A_+$. In the general case, we find a non-vanishing term
\be\label{anomvariation}
\dd S_1  = - {i\over 2\pi} N^a_i Q^b_i A^b_+ {\bar \e} \lambda^a_- + {c.c.},
\ee
which is exactly the way $S_1$ should transform under a superspace gauge transformation with chiral superfield gauge parameter
\be
\Lambda^a = 2i \th^+ {\bar \e} A_+^a. 
\ee
This is exactly the gauge transformation needed to restore Wess-Zumino gauge. Note that $S_2$ given in~\C{s2}\ is also not supersymmetric for the same reason and transforms in a way that precisely cancels the antisymmetric part of~\C{anomvariation}.

What this immediately implies for us is that the target space geometry that emerges from our construction need not be complex because we no longer have classical $(0,2)$ supersymmetry. 
To build string compactifications with space-time supersymmetry, we require models with unbroken $(0,2)$ supersymmetry. In past work on supersymmetry anomalies, it was noted that the one-loop gauge anomaly is accompanied by a corresponding supersymmetry anomaly~\cite{Itoyama:1985ni, Itoyama:1985qi, Hwang:1985tm}. When the gauge anomaly cancels, the supersymmetry anomaly also cancels. This was noted in~\cite{Adams:2006kb}. We therefore expect a quantum $(0,2)$ supersymmetry to exist in all quantum gauge invariant models.  The implications of a quantum $(0,2)$ supersymmetry for the target space are rather mysterious and quite fascinating.

However, there are cases where we do expect complex target spaces. The simplest examples  come from familiar classically gauge-invariant models (there are many interesting models of this type with log interactions). There are even cases which are not classically gauge-invariant but still possess complex target spaces. Examples of this type were constructed in~\cite{Adams:2006kb}, and it is worth describing how they work in our framework. If we make the substitution, 
\be
\Phi^i \, \rightarrow \, e^{Y^i},
\ee
described around equation~\C{defY}, we find non-canonical kinetic terms for the $Y^i$ fields of the form
\be
- e^{Y^i + {\bar Y}^i} | \p_\mu Y^i + i Q_i A_\mu|^2.
\ee
With this choice of kinetic term, there is a coupling of $Y^i$ to the $D$-term proportional to 
\be
Q_i |e^{Y^i}|^2 D
\ee
as well as a term proportional to $N_i Y^i D$, coming from the log interaction expressed in terms of $Y^i$.  This is just a rewriting of the interactions we described earlier in terms of $Y^i$. 

Let us contrast this with the construction of~\cite{Adams:2006kb}\ which involves taking flat kinetic terms for the $Y^i$ fields
\be
- {1\over 2\pi} | \p_\mu Y^i + i Q_i A_\mu|^2.
\ee
This choice corresponds to non-canonical kinetic terms in terms of $\F^i$ fields. With the same FI couplings, this choice gives the following combined coupling of $Y^i$ to $D$:
\be
- {1\over 2\pi}  D\left( Q_i - N_i \right) Y^i. 
\ee
Now setting $Q_i=N_i$ decouples $Y^i$ from $D$. This is something that can never happen with the canonical choice of kinetic terms for $\F^i$. If the $Y^i$ fields decouple from the $D$-term constraint then solving the $D$-term constraint and quotienting by the gauge group gives the standard holomorphic reduction on the remaining fields. This is why the models found in~\cite{Adams:2006kb}\ have complex target spaces, despite the violation of classical gauge invariance. However, models in which the $D$-term constraints are modified by log interactions should possess some more interesting analogue of complexity. It would be interesting to study models that involve both canonical and non-canonical kinetic terms.

\subsection{Comments on the target space geometry}

In components the total bosonic action takes the form,
\be
\L_B = -|\del_\mu \f^i|^2 + j^a{}^\mu A_\mu^a - \hlf \left( \left( \D^{-1}\right)^{ab}\eta^{\mu\nu} - {N_{ i}^{[a}Q_i^{b]}\over2\pi}\e^{\mu\nu}\right) A^a_\mu A^b_\nu + {1\over2\pi}\Th^a F^a - V(\f),
\ee
where
\bea
\Th^a &=& \th^a + N_i^{a} \hat{\th}^i,  \\
V(\f) &=& \sum_a {e_a^2\over2} \left(\sum_i Q_i^a |\f^i|^2 + \sum_i {N_i^{a}\over2\pi}\log|\f^i| - r^a \right)^2.
\eea
In the expression for $\Th^a$, we have used $\hat{\th}^i = \Im(\log(\f^i))$ for the phases of the $\f^i$ fields.

To extract the classical target space metric, we need to integrate out the gauge field. As observed in~\cite{Adams:2009tt}, this is complicated by the fact that the classical action is not gauge invariant. One way to remedy this is to suppose a transformation law for $\th^a$
\be
\th^a \mapsto \th^a + {1\over 2}\cA^{ab}\La^b.
\ee
Alternatively, we can incorporate the effect of the anomaly by adding an appropriate term to the one-loop effective action. By considering the descent relations, we see that we cannot write this term as a coupling directly on the world-sheet, but we can write it formally as part of an action in one higher dimension. As in the case of a WZW term, we consider a $3$-manifold $\cC$ whose boundary $\del \cC=\Sigma$ is the string world-sheet. Let $t$ coordinatize the extra direction. We need to extend our gauge fields into the interior of $\cC$ so let
\be
\tilde{A}^a = \tilde{A}^a(x,t) \qquad {\rm with} \qquad \tilde{A}^a(x,0) = A^a(x).
\ee
The anomaly can now  effectively be written as
\be
S_{one-loop} =  {\cA^{ab}\over 4\pi} \int_\cC  \tilde{A}^a\d\tilde{A}^b. 
\ee
We have made a choice of orientation so that $\int_\cC d(\cdot) = - \int_\Sigma$. In fact, our $AV$ coupling also takes this form because of its anti-symmetry:
\be
{1\over 4\pi}{N_i^{[a}Q_i^{b]}}\int_\Sigma A^a\wedge A^b ={1 \over 2\pi} {N_i^{[a}Q_i^{b]}} \int_\cC \tilde{A}^a \d\tilde{A}^b.
\ee
Since we require $Q^{(a}_i N_i^{b)} = -{1\over 2}\cA^{ab} $ for a consistent theory, it  makes sense to combine these into a single Chern-Simons term
\be
S_{CS} = - {N_{ i}^{b} Q_i^{a} \over 2\pi} \int_\cC \tilde{A}^a \d\tilde{A}^b.
\ee
The bosonic action then takes the form,
\be
S_B = \int_\Sigma\d^2x\, \Big(-|D_\mu\f^i|^2 + {1\over2\pi}\Th^aF^a -V(\f)\Big) - {N_{ i}^{b} Q_i^a\over2\pi} \int_\cC \tilde{A}^a\d\tilde{A}^b.
\ee
The IR $B$-field will be given by $B = \Th^aF^a$, which is neither closed nor gauge invariant. We can lift $B$ naturally to $\cC$ and combine it with the Chern-Simons terms to get the gauge-invariant field strength $H$:
\be
S_H = {1\over2\pi}\int_\cC \f^*(H) = - {1\over2\pi} \int_\cC \big(\d\Th^a + N_{ i}^{a} Q_i^b\tilde{A}^b\big)\tilde{F}^a.
\ee
The flux is therefore given by the quantized expression $H = - N_i^{a}(\d \hat{\th}^i + Q_i^b A^b)\wedge F^a$. While gauge invariant, this $H$ is not closed. Taking its curl, we find the modified heterotic Bianchi identity
\be \label{dHequation}
\d H = - N_{i}^{a} Q_i^b F^b\wedge F^a = {1\over 2}\cA^{ab} F^a\wedge F^b = {\rm ch}_2({\cal E}) - {\rm ch}_2(X) 
\ee
where $ {\cal E}$ is the gauge bundle determined by the left-moving fermions, and we have used the symmetry of $F\wedge F$ to project onto $-N_{i}^{(a}Q_i^{b)}={1\over 2}\cA^{ab}$. The last equality in~\C{dHequation}\ is the non-linear sigma model interpretation of the one-loop anomaly.  Note that if our space includes loci where log terms can become singular then $d\hat{\th}$ is not closed and gives an additional
delta-function contribution to $dH$ corresponding to (anti-)NS5-brane sources denoted $[B]$ in~\C{bianchi}.

Including the Chern-Simons term gives a gauge-invariant one-loop effective action. It should now be possible to integrate out the gauge-fields $A^a$ following the discussion in section~\ref{classicalgeometry}\ to find expressions for the metric and $B$-fields for these spaces. This would help shed light on how the $H$-flux is supported in the geometry. 

We should also stress that  generically the $r^a$ parameters do not correspond to moduli. The two-cycles whose volumes they apparently measure are trivialized if the associated $\Th^a$ circle bundle is non-trivial. Essentially, the flux removes these cycles from the geometry. 

\subsection{Compact non-conformal examples}
\label{compactex}

Let us consider the case of one $U(1)$ initially. The $D$-term constraint becomes
\be\label{simpleD}
\sum_i Q_i |\phi^i|^2 + {N_i\over 2\pi} \log |\phi^i| = r.
\ee
For the simplest compact model, let us assume all $Q_i>0$ and $r>0$. For large fields $|\phi^i|$, the log terms are irrelevant and we approximate weighted projective space. The dangerous region is when a $\phi^i$ with non-zero $N_i$ becomes small. However, if all $N_i\leq 0$ then this region is excluded.  $N^i$ negative corresponds to an anomaly contribution from NS5-brane flux or a gauge instanton. The flux bounds us away from the sources where one or more $\phi^i$ vanish. 

For anti-NS5-brane flux, where at least one $N^i$ is positive, the solution for~\C{simpleD}\ becomes non-compact and develops a throat near the singularity. For this case, we see the brane source and the metric is dominated by the log terms. 

We can generalize this construction to many $U(1)$ fields and constraints:
\be\label{manyD}
\sum_i Q_i^a |\phi^i|^2 + {N_i^a \over 2\pi} \log |\phi^i| = r^a.
\ee
In this case, it need not be the case that all $N_i^a$ are negative. As a simple example with $U(1)\times U(1)$ gauge group, take the charge matrix
\be
Q_i^a = \left(\begin{array}{cccccccc} 1 & 1 & \ldots & 1& 0 & 0 &\ldots & 0 \\ 0 & 0 & \ldots & 0 & 1 & 1 & \ldots & 1\end{array}\right)
\ee
where the first block has length $n$ and the second length $m$. Assume $m \geq n$. Now add a set of $n+m$ left-moving fermions with
\be
Q_m^\a = \left(\begin{array}{cccccccc} 0 & 0 & \ldots & 0& 0 & 0 &\ldots & 0 \\  1& 1 & \ldots & 1 & 1 & 1 & \ldots & 1\end{array}\right).
\ee
Take $N_i^{2} = - N_i^{1} = {1\over 2}$ for $i=1,\ldots, 2n$ and $0$ otherwise.  It is easy to check that the quantum anomaly gives
\be
\cA^{ab} = \sum_i Q_i^a Q_i^b - Q_m^\a Q_m^\b =\left(\begin{array}{cc} n & 0 \\ 0 & m \end{array}\right) -\left(\begin{array}{cc} 0 & 0 \\ 0 & n+m \end{array}\right)  = \left(\begin{array}{cc} n & 0 \\ 0 & -n \end{array}\right)
\ee
while
\be
2 N_i^a Q_i^b = \left(\begin{array}{cc} -n & -n \\ n & n \end{array}\right).
\ee
We see that the diagonal part is canceled by $\cA^{ab}$ while the off-diagonal part is canceled by the $AV$ coupling.

It is instructive to examine the $D$-term constraints of this model:
\bea
&& \sum_{i=1}^n |\f^i|^2 + {1\over 4\pi}\sum_{i=1}^{2n} \log|\f^i| = r_1 \\
&& \sum_{i=n+1}^{n+m} |\f^i|^2 - {1\over 4\pi}\sum_{i=1}^{2n} \log|\f^i| = r_2.
\eea
Despite the presence of the (unbounded) log interactions, the vacuum manifold is nonetheless compact. To see this, consider the sum of the $D$-terms
\be
\sum_{i=1}^{m+n} |\f^i|^2 = r_1+ r_2 \equiv r_+,
\ee
which implies the space is compact. In particular, after quotienting by $U(1)$ it is $\PP^{n+m-1}$.  The difference of the $D$-terms,
\be
\sum_{i=1}^n |\f^i|^2 - \sum_{i=n+1}^{n+m} |\f^i|^2 +{1\over 2\pi} \sum_{i=1}^{2n} \log|\f^i| = r_1 -r_2 \equiv r_-,
\ee
carves out a (real) hypersurface in this compact projective space, so the final space is ultimately a smooth compact surface. Further quotienting by $U(1)$ would usually give a complex target space. As discussed in section~\ref{susyanomaly}, we expect the target space to be generically non-complex.\footnote{ For example, consider a single $U(1)$ with $D$-term:
\be
|\f_1|^2 + |\f_2|^2 + |\f_3|^2 - {3\over 2\pi} \log |\f_1| = r.  
\ee
The field $\f_1$ can never vanish so we can fix the $U(1)$ action by choosing $\f_1$ real and positive. The solution set for $\f_1$ is then an interval. At the ends of the interval $|\f_2|^2 + |\f_3|^2$ vanish. The geometry is therefore an $S^3$ fibered over the $\f_1$ interval with the $S^3$ going to zero size at the ends of the interval. This space is $S^4$, which admits no complex or almost complex structure. A detailed study of this and related models will appear elsewhere in collaboration with Mark Stern.}

To get a better feel for the structure here, let us take the very simplified case of $n=m=1$ for which the vacuum structure is still interesting. In this case, 
\be
|\f_1|^2 + |\f_2|^2 = r_+,
\ee
where $\f_1$ solves
\be \label{trans}
|\f_1|^2(r_+-|\f_1|^2)e^{ 8\pi |\f_1|^2} = e^{8\pi r_1}.
\ee
Although generally  it is not possible to invert this transcendental equation, it is easy to extract  key features of the solution set. First, $r_+>0$ and $|\f_1|^2 \leq r_+$ for a solution.  The left hand side of~\C{trans}\ is therefore positive with a maximum. The function on the left hand side is actually very sharply peaked. To get a feel for the shape,  figure~$1$ plots $|\f_1|^2(1-|\f_1|^2)e^{ 4\pi |\f_1|^2}$ versus $|\f_1|^2$, which is less sharply peaked than~\C{trans}\ permitting the maximum to be visible on the graph. 

\begin{figure}[ht]
\begin{center}
\[
\mbox{\begin{picture}(300,200)(0,20)
\includegraphics[scale=0.5]{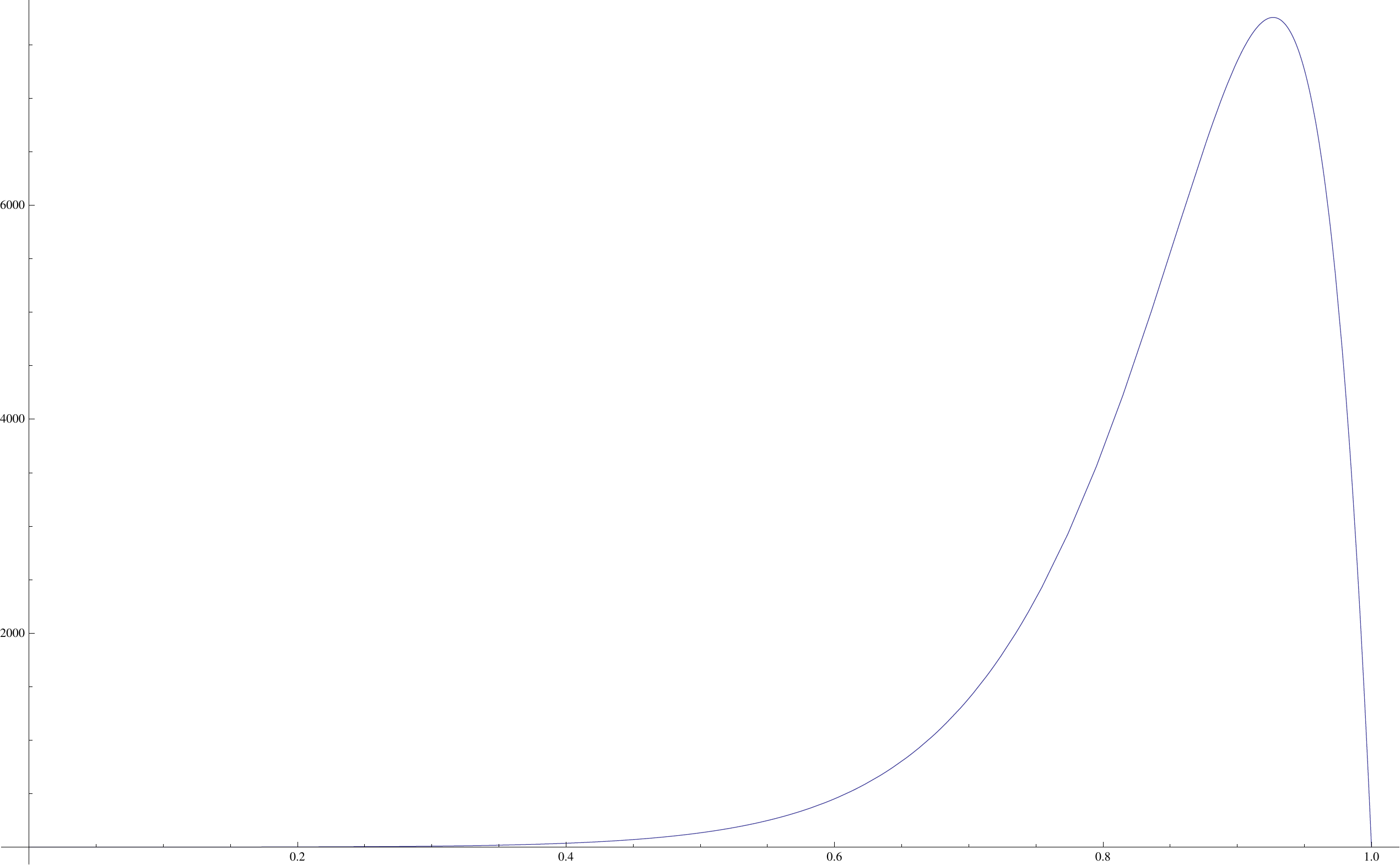}
\end{picture}}
\] 
\vskip 0.2 in \caption{\it A plot of $|\f_1|^2(1-|\f_1|^2)e^{ 4 \pi |\f_1|^2}$ against $ |\f_1|^2$. The case of~\C{trans}\ with $e^{ 8 \pi |\f_1|^2}$ rather than $e^{ 4 \pi |\f_1|^2}$ is qualitatively similar but exponentially more sharply peaked. There are two solutions for $|\f_1|^2$  except for the maximum value, as long as $r_1$ is sufficiently small.}
\end{center}
\end{figure}

For values of $r_1$ below  the maximum, there are two values of $|\f_1|^2$ solving~\C{trans}. 
This is already quite different from weighted projective space. In general $r_1$ need not be positive! For example, taking $r_1 \rightarrow -\infty$ while holding fixed $r_+>0$ gives two solutions: $(|\f_1|^2 = r_+, |\f_2|^2=0)$ and $(|\f_2|^2 = r_+, |\f_1|^2=0)$. 
The maximum value for the left hand side of~\C{trans}\ is attained when
\be
|\f_1|^2 =  \left( {1\over 2} \, r_+  - {1\over 8\pi} + {1\over 8 \pi}\sqrt{1 + 16\pi^2 (r_+)^2} \right).
\ee
For this value, there is a unique choice of $r_1$ permitting a vacuum solution; for larger values of $r_1$, there are no vacuum solutions. For $r_+$ very small, only negative choices for $r_1$ admit vacuum solutions. The crossover occurs for $r_+ \sim 0.23$. Even from this simple example, it is clear that the structure of the moduli space is going to be quite fascinating as a function of the $r$ parameters.


\subsection{Non-compact  examples}
\label{noncptex}

The simplest example of a non-compact space is a model with all positive charges but with some positive $N_i^a$; for example, a case like~\C{simpleD}\ with a single $U(1)$ factor. These models are fascinating because the log terms {\it add} to the quantum anomaly like a $\Tr (R\wedge R)$ contribution. To cancel the anomaly, we introduce a left-moving gauge bundle with appropriate charges. There are many ways to do this, and the degeneracy of such solutions grows very quickly. The picture for this case is a collection of anti-NS5-branes assembled in the geometry contributing to the tadpole. 

Total charge neutrality is ensured by adding sufficient numbers of instantons and NS5-branes. The existence of throats in the geometry does not imply these models are necessarily non-compact as space-time backgrounds, but rather that we have to introduce sources. We do suspect string perturbation theory breaks down, though that should be analyzed more carefully in models with multiple $U(1)$ factors. One of the puzzling aspects of these models is whether they are space-time supersymmetric. We plan to examine these models in detail elsewhere. 

We can also generalize the usual construction of non-compact toric  spaces where some fields have negative charges. This includes the case of non-compact Calabi-Yau spaces for which the charges must sum to zero for each gauge factor. Let us denote the negatively charged fields by $p^n$ with charges $- Q_n^a$. Take the case of one $p$-field and a single $U(1)$ with $D$-term constraint
\be\label{withp}
\sum_i Q_i |\phi^i|^2 + {N_i \over 2 \pi} \log |\phi^i| = r + Q_p |p|^2.
\ee
If all $N_i$ were zero, this describes the total space of the line bundle $O(-Q_p) \rightarrow W\PP_{\{Q_i\}}$ over weighted projective space. In this case, we could take $r$ negative and consider a point in the moduli space where $\phi^i=0, \, p \neq 0$. At such an orbifold point, the $U(1)$ gauge group is broken to a discrete group $\Z_{Q_p}$ for $Q_p>1$. 

However, if any $N_i$ are non-vanishing and negative then the associated $\phi^i$ can never be taken to zero and this orbifold point is removed from the moduli space if $\phi^i$ is charge $1$. Should $\phi^i$ have charge greater than $1$ then the unbroken discrete group is $\Z_{\rm gcd(Q_i, Q_p)}$ with obvious generalizations. The space can again be viewed as the total space of a line bundle but over the non-K\"ahler generalization of weighted projective space described in the preceding section. 

We can generalize this construction further by considering log terms for the $p$-fields. Let us return to the case of the conifold considered in section~\ref{conifoldsec}. Generalize the model as follows:
\bea\label{ppluslog}
 |\phi^1|^2 +  |\phi^2|^2 +  {N_1\over 2 \pi} \log |\phi^1| +  {N_2\over 2\pi} \log |\phi^2| &=& r + |p^1|^2 + |p^2|^2 \cr && + {M_1\over 2 \pi} \log |p^1|+ {M_2 \over 2 \pi} \log |p^2|.
\eea
If we choose $N_1+N_2 - M_1 - M_2=0$ then the gauge anomaly is unchanged. This is really a case of gauge invariant $f$ described in section~\ref{gaugeinvariant}\ but with a log of a product of monomials. For other choices of $(N,M)$, the anomaly changes but can typically  be canceled by an appropriate choice of left-moving gauge bundle. 

The geometry now has collections of throats when some $N$ or $M$ are non-vanishing. In the usual case where $N=M=0$, there is a symmetry taking $ r\rightarrow -r$ and exchanging the roles of the $\phi$ and $p$-fields corresponding to a flop. There is an analogous transition here from $r$ to $-r$ but with an additional exchange of $N$ and $M$. It would be interesting to understand this transition geometrically.

\subsection{Superpotentials and Intersections}

Superpotential couplings in $(0,2)$ theories require left-moving fermions. Consider a possibly charged left-moving fermion $\G$ with
\be
\bar{\mathfrak{D}}_+\G = \sqrt{2} E(\F).
\ee
We can introduce superpotential couplings
\be\label{super}
S_J = -{1\over \sqrt{2}}\int\d^2x\d\th^+\, \G \cdot J(\F) + c.c.,
\ee
supersymmetric if $E\cdot J=0$, which give a bosonic potential
\be
V = |E|^2 + |J|^2. 
\ee
For the moment, let us set $E=0$. Generic choices for $J$ have moduli spaces that consist of points so we want to make non-generic choices to find interesting geometric and non-geometric phases. Introducing any superpotential certainly changes the determination of $R$-symmetries and the conditions for conformal invariance.  The main assumption we will make is that the log superpotential interactions do not significantly alter the usual conformality arguments. This assumption is based to a large extent on space-time expectations.  When the singular regions of the log interactions are excluded from the moduli space, this seems quite reasonable, but it really should be checked carefully.     

For convenience, let us split our fields into $\F^i$ and $P^n$ with no log interactions and  positive and negative charges $(Q_i, -Q_n)$ respectively, $\vF^j$ and $\vP^m$ with non-vanishing log interactions and  charges $(Q_j, -Q_m)$.

\subsubsection{A single $U(1)$ and a single $P$-field}

We expect the IR geometry for a conformal target space $\M$  to have vanishing $c_1$ so let us take a superpotential
\be \label{hyper}
S_J = -{1\over \sqrt{2}}\int\d^2x\d\th^+\, {\hat \G} W(\F, \vF) + \G^\a P J_\a (\F, \vF) + c.c.
\ee
with ${\hat \G}$ of charge $Q_\G$. We choose
\be
Q_{\hat \G} + \sum_i Q_i + \sum_j Q_j =0
\ee
to ensure $c_1(\M)=0$. It is also usual to take $c_1({\cal E})=0$ for the IR bundle ${\cal E}$.
It might be possible to weaken this condition but let us impose it by requiring
\be
Q_P = \sum_\a Q_\a. 
\ee
This model is not conformal unless 
\be \label{uvconformal}
Q_P = \sum_i Q_i + \sum_j Q_j
\ee
which is generally not true except for particularly nice models like those with $(2,2)$ supersymmetry. Our situation is no different than conventional $(0,2)$ models where spectator fields are included to ensure~\C{uvconformal}\ is satisfied~\cite{Distler:1995mi}. Namely, add a spectator chiral superfield $S$ with charge 
\be
Q_S =  Q_P - \sum_i Q_i - \sum_j Q_j 
\ee
and left-moving partner $\G_S$ with opposite charge together with the superpotential interaction $ \int \d\th^+\, \G_S S$. Lastly, we insist on gauge anomaly cancelation but these superpotential interactions do not affect our earlier discussion.

In the usual argument for a geometric phase, we want $P=0$ for $r>0$. If we were to replace $P$ by $\vP$ in~\C{hyper}, the $D$-term constraints would prevent this possibility.  To engineer a geometric phase, pick a transverse $W$ defining a non-degenerate hypersurface $W=0$ in the ambient non-K\"ahler space. Choose $J_\a$ such that at least one is non-vanishing on this hypersurface thereby forcing $P=0$. That is easy to arrange since $\vF$ never vanish. For these models, $r$ need not be positive in this geometric phase! The range of $r$ admitting a geometric interpretation will depend on the choice of charges and $N$ coefficients.     

If we take $r$  negative in theories without log interactions, we would encounter a Landau-Ginzburg phase. As discussed in section~\ref{intro}, our $r$ parameter does not correspond to a modulus in the presence of log interactions. We can still examine how the physics changes when $r$ lies in different ranges.  For $r$ sufficiently negative, $P$ must be non-vanishing to satisfy~\C{withp}. We must then satisfy $W = J_\a =0$. Since the $\vF^j$ never vanish, these constraints need not force $\F^i=0$ and the gauge group is typically still broken. 

For superpotentials of type~\C{hyper}, there is typically a non-compact moduli space for sufficiently negative $r$ where $|P|$ and $|\vF|$ become large. We can engineer better behaved models in this parameter region by allowing some $J$ couplings in~\C{super}\ to become more interesting functions of $P$. For example, let $ x_k$ denote the gauge-invariant monomials constructed from $P$ and $|\vF|$. Introduce a superpotential
\be
\int \d\th^+\, \G  \cdot g_1(\vF) g_2(x_k)
\ee 
with $g_1$ polynomial and $g_2$ some reasonable function with  zero at $P=0$. With $P \neq 0$ in the $r$ sufficiently negative region, some or all of the $\vF^j$ are fixed at roots of $g_2$. A residual discrete gauge group is possible in this case. If the remaining constraints $J^i=0$ and $W=0$ force $\F^i=0$, we would have a Landau-Ginzburg phase. However, such models typically have no nice geometric phase. Another class of models with no geometric phase would involve potentials with $\vP$ fields. 

\subsubsection{More general models}

It should already be clear that there is a fairly complex space of models possible in this framework. There are straightforward generalizations to complete intersections obtained by including many ${\hat \G}^\e$ with label $\e$. Allowing multiple gauge groups is also straightforward. A standard superpotential for the geometric phase would take the form
\be \label{manyhyper}
S_J = -{1\over \sqrt{2}}\int\d^2x\d\th^+\, {\hat \G}^\e W_\e(\F, \vF) + \G^\a P_l J_\a^l (\F, \vF) + c.c.
\ee
with charge constraints
\bea
\sum_\e Q^a_{ {\hat \G}_\e} + \sum_i Q^a_i + \sum_j Q^a_j &=& 0, \cr
\sum_l Q_{P_l}^a - \sum_\a Q_\a^a &=& 0. 
\eea
coupled with gauge anomaly cancelation and possibly spectators for conformal invariance. With multiple $U(1)$ factors, we do expect to find hybrid phases with Landau-Ginzburg components when we vary a combination of $r$ parameters that actually corresponds to a modulus.

\subsection*{Acknowledgements}

It is our pleasure to thank Allan Adams, Ilarion Melnikov and Stefan Groot Nibbelink for discussions. We would also like to thank the organizers and participants of the
 ``Topological Heterotic Strings and (0,2) Mirror Symmetry Workshop" in Vienna hosted by the Erwin Schr\"odinger Institute.
 \vskip 0.1in

C.~Q. is supported in part by NSF Grant No.~PHY-0758029 and by an NSERC PGS-D research scholarship. S.~S. is supported in part by
NSF Grant No.~PHY-0758029 and NSF Grant No.~0529954.

\appendix
\section{The Chiral Gauge Anomaly}
\label{anomalyapp}

We are interested in computing  the one-loop gauge anomaly paying close attention to the overall sign of the contribution. We will try to be as general as possible and state explicitly any assumptions we are making about our conventions. Let us begin by considering two-dimensional gamma matrices in a chiral basis but with otherwise arbitrary (complex) coefficients. We also leave the overall sign of the Minkowski metric undetermined. Let
\be
\g^0 = \mat{cc}{0 & \a_1 \\ \a_1 & 0},\qquad \g^1 = \mat{cc}{0 & \a_3 \\ \a_4 & 0},\qquad \eta_{\mu\nu} = \mat{cc}{-s & 0 \\ 0 & s}, 
\ee
where $\a_i$ are complex phases with $|\a_i|^2=1$ and $s=\pm1$. As long as we insist that the $\g^\m$ satisfy the Clifford algebra
\be
\{\g^\m,\g^\n\} = 2\eta^{\m\n}
\ee
then the gamma matrices are determined up to a choice of complex phase $\a$ and two real phases $s,c=\pm1$
\be
\g^0 = \mat{cc}{0 & \a \\ -s\a^* & 0},\qquad \g^1 = \mat{cc}{0 & sc\a \\ c\a^* & 0},\qquad \g^5 = \g^0\g^1 = \mat{cc}{c & 0 \\ 0 & -c}.
\ee
This fixes the Dirac operator to be
\be
i\cD\!\!\!\!\!\;/ = i\mat{cc}{0 & \a(\cD_0+sc \cD_1) \\ -\a^*s(\cD_0-sc \cD_1) & 0}.
\ee
We will write our two-component spinors and their Hermitian conjugates as
\be
\j = \mat{c}{\j_1 \\ \j_2},\qquad \bar{\j} = \rho \j^\dagger\g^0,
\ee
where usually $\rho=1$ or $i$ depending on a choice of convention and whether $\psi$ is real or complex. The Dirac action then splits into two chiral pieces
\be
\L_F = \s\bar{\j}\cD\!\!\!\!\!\;/\j = -s\s\rho\Big(\j_1^\dagger (\cD_0-sc \cD_1)\j_1 + \j_2^\dagger(\cD_0+sc \cD_1)\j_2\Big)
\ee
where $\s$ is another convention/representation-dependent phase which ensures the total action is real. Luckily, our life simplifies a little here since $s\s\rho=-i$ in any convention. We now define the chiral projections:
\be
\j_\pm = \hlf(1\pm s\g^5)\j.
\ee
These have the property that $\g^5\j_\pm = \pm s\j_\pm$. More importantly, when $sc=\pm1$ we have $\j_1=\j_{\mp}$ and $\j_2 = \j_\pm$. The advantage of this definition is that the chiral action takes the form
\be
\L_F = i\Big( \j_+^\dagger \cD_-\j_+ + \j^\dagger_- \cD_+ \j_-\Big)
\ee
where $\cD_\pm = \cD_0 \pm \cD_1$ for any choice of $s,c$ and $\a$. We will refer to $\j_+$ ($\j_-$) as right- (left-) movers. We stress that this entire discussion simply defines what we mean by the labels $\j_\pm$ and does not adhere to any single convention.

Now that we have set things up in as convention independent of a form as we could, we may now go ahead an compute the anomaly. We will follow the method of Fujikawa~\cite{Fujikawa:1979ay}, and study the transformation properties of the path-integral measure under global chiral transformations. Using a two-component notation, the spinors transform by
\be
\j_i \mapsto \exp\Big({{i\over2}(1\pm s\g^5)\a^a Q^a_i}\Big)\j_i ,\qquad \bar{\j}_i \mapsto \bar{\j}_i \exp\Big({{i\over2}(-1\pm s\g^5)\a^a Q^a_i}\Big). 
\ee
The measure therefore changes by
\bea
\prod_i D\j_i D\bar{\j}_i &\mapsto& \prod_i D\j_i D\bar{\j}_i  \det{}^{-1}\left(\exp\Big({{i\over2}(1\pm s\g^5)\a^a Q^a_i}\Big)\exp\Big({{i\over2}(-1\pm s\g^5)\a^a Q^a_i}\Big)\right) \non\\
&=& \prod_i D\j_i D\bar{\j}_i \exp\Big(\mp is\Tr\big(\g^5 Q^a_i\a^a\big) \Big).
\eea
This means that the action shifts by
\be
\dd S = \mp s \Tr(\g^5 Q^a_i\a^a) = \mp s\sum_i Q^a_i \int\d^2x\ \a^a \Big[\sum_n \f_{ni}^\dagger(x) \g^5 \f_{ni}(x) \Big]. 
\ee
To compute the trace of $\g^5$, we have expanded in a complete eigenbasis of the Dirac operator for fields with charges $Q^a_i$
\be
i\cD\!\!\!\!\!\;/ \f_{ni} = \la_n \f_{ni}.
\ee
To regulate this trace, we perform the usual trick of introducing a convergence factor $e^{-s\la_n^2/M^2}$ where the factor of $-s$ is needed to ensure convergence for large values of $\la_n$. To see this, note that
\be
(i\cD\!\!\!\!\!\;/)^2 = -\cD^2 - iQ^a_i F^a_{01}\g^5
\ee
which approaches $-\del^2$ at large momenta with fixed background $A$. This means $\la_n^2$ approaches $k^2= -s(k_0^2 -k_1^2) \simeq s k_E^2$ which is negative definite for Euclidean momenta ($k^0=ik_E^0$) only when multiplied by $-s$. Now we compute the trace in the usual manner:
\bea
\lim_{M\rightarrow\infty} \sum_n \f_{ni}(x)^\dagger \g^5 e^{-s(i\cD\!\!\!\!\!\;/)^2/M^2}\f_{ni}(x) 
&=& \lim_{M\rightarrow\infty} \tr \langle x |\g^5 e^{-s(-\cD^2-i\g^5Q^a_iF^a_{01})/M^2} |x\rangle \\
&=& \lim_{M\rightarrow\infty} \langle x |e^{s\del^2/M^2} |x\rangle \tr \left[\g^5 (1+is \g^5Q^a_iF^a_{01})/M^2)\right] +\ldots\non\\
&=& \lim_{M\rightarrow\infty}\left({iM^2\over4\pi}\right)\left({2isQ^a_i F^a_{01}\over M^2} \right) +\ldots \non\\
&=& -{s\over2\pi}Q^a_iF^a_{01}. \non
\eea
Therefore the global chiral anomaly changes the action by
\be
\dd_{\rm global} S = {\cA^{ab}\over2\pi}\int\d^2x\ \a^a F^b_{01}
\ee
where the anomaly coefficient $\cA^{ab}$ is determined by the charges $Q^a_i$ of the right-movers, $\j^i_+$, and the charges $Q^a_\a$ of the left-movers, $\j^\a_-$,
\be
\cA^{ab} = \sum_i Q^a_i Q^b_i - \sum_\a Q^a_\a Q^b_\a.
\ee
As a consistency check, we see that all the convention/representation dependent coefficients $\a,s,c,\rho,\s$ drop out of the final result as they should. 

The gauge anomaly is closely related to the global chiral anomaly derived here. The primary difference is a subtle factor of $2$ giving: 
\be
\dd_{\rm gauge} S = {\cA^{ab}\over4\pi}\int\d^2x\ \a^a(x) F^b_{01}.
\ee
A careful derivation can be found in~\cite{Melnikov:2012nm}.

\newpage

\ifx\undefined\bysame
\newcommand{\bysame}{\leavevmode\hbox to3em{\hrulefill}\,}
\fi




\end{document}